# Achieving 5-angstrom-resolution diffraction contrast with an uncorrected electron probe and nanosized defect characterizations

Yao Li[1,2], Austin C. Houston[3], Ziang Yu[4], Zehui Qi[1], Siwei Chen[1], Yajie Zhao[1], Sung Joo Kim[5], Steven J. Zinkle[1,3,6*], Gerd Duscher[3], Blas P. Uberuaga[7], Benjamin K. Derby[2,8]

[1] Department of Nuclear Engineering, University of Tennessee, Knoxville, TN 37996, USA

[2] Center for Integrated Nanotechnologies, Los Alamos National Laboratory, Los Alamos, NM 87545, USA

[3] Department of Material Science and Engineering, University of Tennessee, Knoxville, TN 37996, USA

[4] Department of Nuclear Engineering & Engineering Physics, University of Wisconsin, Madison, WI 53706 USA

[5] Electron Microscopy Center, University of Tennessee, Knoxville, TN 37996, USA

[6] Materials Science and Technology Division, Oak Ridge National Laboratory, Oak Ridge, TN 37830, USA

[7] Materials Science and Technology Division, Los Alamos National Laboratory, Los Alamos, NM 87545, USA

[8] Woodward, Inc., Fort Collins, CO 80524, USA

*Corresponding author

Email: S. Zinkle (szinkle@utk.edu)

Co-author email addresses:

Y. Li (yli166research@outlook.com)A. Houston (ahoust17@vols.utk.edu); Z. Yu (ziangyu1994@gmail.com); Z. Qi(zqi1@vols.utk.edu); Siwei Chen (schen83@utk.edu); Y. Zhao (yzhao65@utk.edu); S. Kim (skim172@utk.edu); S. Zinkle (szinkle@utk.edu); G. Duscher (gduscher@utk.edu); B. Uberuaga (blas@lanl.gov); B. Derby (benjamin.derby@icloud.com)

# Abstract

Atomic-resolution micrographs remain limited in application due to small field of view and low numbers of recorded defects. Defect imaging with reliable statistics in transmission electron microscopy (TEM) relies heavily on diffraction contrast. However, with 200-300 kV electrons, low-curvature Ewald sphere limits diffraction resolution at ~5 nm. Here, we imaged defects with a forbidden condition in TEM and improved the resolution in diffraction contrast from 5 to 0.5 nm. The superiority of our 'equal-s' method is demonstrated by the discovery of a novel phase and dislocation loop structure in steels, which have not been previously reported experimentally or theoretically. Our ‘equal-s’ can largely replace the classical ‘two-beam’ condition for defect study.

# Introduction

In fields such as structural biology, physics, chemistry, and materials science, conventional and scanning transmission electron microscopy (CTEM and STEM) have been the "gold standard" for studying micro- and nano defects. These defects are clusters of secondary elements or atoms of major elements displaced from original positions in crystals. Although “defects” implies a detrimental impact, they are paramount for advancement of science and engineering such as interface-induced superconductivity (*1*), dislocation-assisted high-ductility, high-strength alloy (*2*), precipitate-enhanced high-temperature application (*3*), and creep-resistant crystals by stable grain boundaries (*4*). Analyzing these tiny defects can be assisted by higher spatial resolution. Thanks to the advancement of spherical aberration ($C_s$) corrections, atomic-resolution can be routinely achieved, especially in STEM. With a state-of-the-art algorithm, atomic-level resolution has approached its physical limit of ~50 pm (*5*). However, atomic-resolution C/STEM is inherently limited by on-zone conditions as a prerequisite. Hence, three-dimensional defects can only be observed in limited directions without resolution degradation, leading to incomplete three-dimensional structural information. Additionally, atomic-resolution images have a tiny field of view, resulting in a low number of defects available for analysis and, consequently, poor statistics. Because of these two inconveniences, atomic-resolution imaging typically is reserved for providing finest details after abundant information has been gathered, primarily through diffraction contrast mechanisms.

Since the 1960s, diffraction contrast under two-beam approximation has been the dominant condition for defect imaging in TEM. The classical two-beam approach can project images of defects called dislocations whose physical widths are ~ 0.2 nm as features with width up to ~20 nm in micrographs when dynamical effect is fully satisfied (excitation error s = 0) (*6*). Thus, if dislocations are closely spaced, microscopists cannot discern individual dislocations in micrographs due to poor resolution. To improve resolution in diffraction contrast, Cockayne developed weak-beam-dark-field (WBDF) TEM with 1-nm resolution demanding $s > 0.2\ nm^{-1}$ (*7*). Unfortunately, in modern TEMs at 200-300 kV forming low-curvature Ewald’s sphere, 1-nm resolution requires ~ $6^{th}$ or higher order reflections to be excited (*8*), which is practically impossible and the classical (g,3g) WBDF can only achieve ~5 nm resolution on important metals, e.g. Ni- or Fe-based alloys. Thus, the electron microscopy characterization gap between atomic-resolution approach and classical two-beam diffraction techniques has left nanosized defects and the internal fine structure of larger defects poorly characterized and corresponding theories unconfirmed. Crystallographic information in this spatial domain such as concentrated or nanosized defects in materials is virtually nonexistent. However, this knowledge is critical. For instance, accumulated high-density dislocations are commonly observed in mechanical fatigue (*9*). Interestingly, concentrated dislocations, introduced by mechanical processing (*2, 10*) or novel additive manufacturing (*11, 12*), also pave the road for alloys with superior mechanical properties. These similar structures yield paradoxical outcomes, encompassing both detrimental (fatigue) and beneficial (strengthening with high ductility) consequences, the underlying causes of which remain inadequately understood. Moreover, the missing crystallographic information of nano-loops from experiments prevents validation of advanced atomistic simulations. For example, over a dozen models regarding **<100>** dislocation loop production have been proposed in Fe (*13*), yet the underlying mechanism(s) remain a subject of heated debate, leaving a critical gap in our knowledge. From micron-length concentrated dislocations with nanometer spacing to embryonic defect clusters close to 1nm in size, microscopists have no tools to spatially resolve them. Thus, experimentally determining defect structures via high-resolution diffraction contrast is critically needed for validating current models, proposing new theories, and engineering advanced materials for extreme conditions.

To overcome this microscopy nanoscale characterization gap, we have developed a novel approach. Traditionally, to minimize image degradation from inelastically scattered electrons, microscopists tend to form images slightly away from the exact Bragg conditions with positive excitation error (s > 0). However, STEM micrograph formation is virtually free from inelastically scattered electrons, enabling defect imaging with negative s. Our method of STEM "equal-s" imaging (Fig. 1) improves diffraction-contrast resolution from 5 nm to 0.5 nm. Using the enhanced capabilities of our new characterization method, we report two new findings in ion-irradiated FeCr alloys: (1) formation of previously undiscovered Body-Centered Tetragonal (BCT) precipitates in Body-Centered-Cubic (BCC) FeCr, and (2) initial nucleation of nanoscale non-edge-type dislocation loops (different from classical expectations), offering unprecedented insight into embryonic defect cluster formation, and enabling direct comparison with atomistic models.

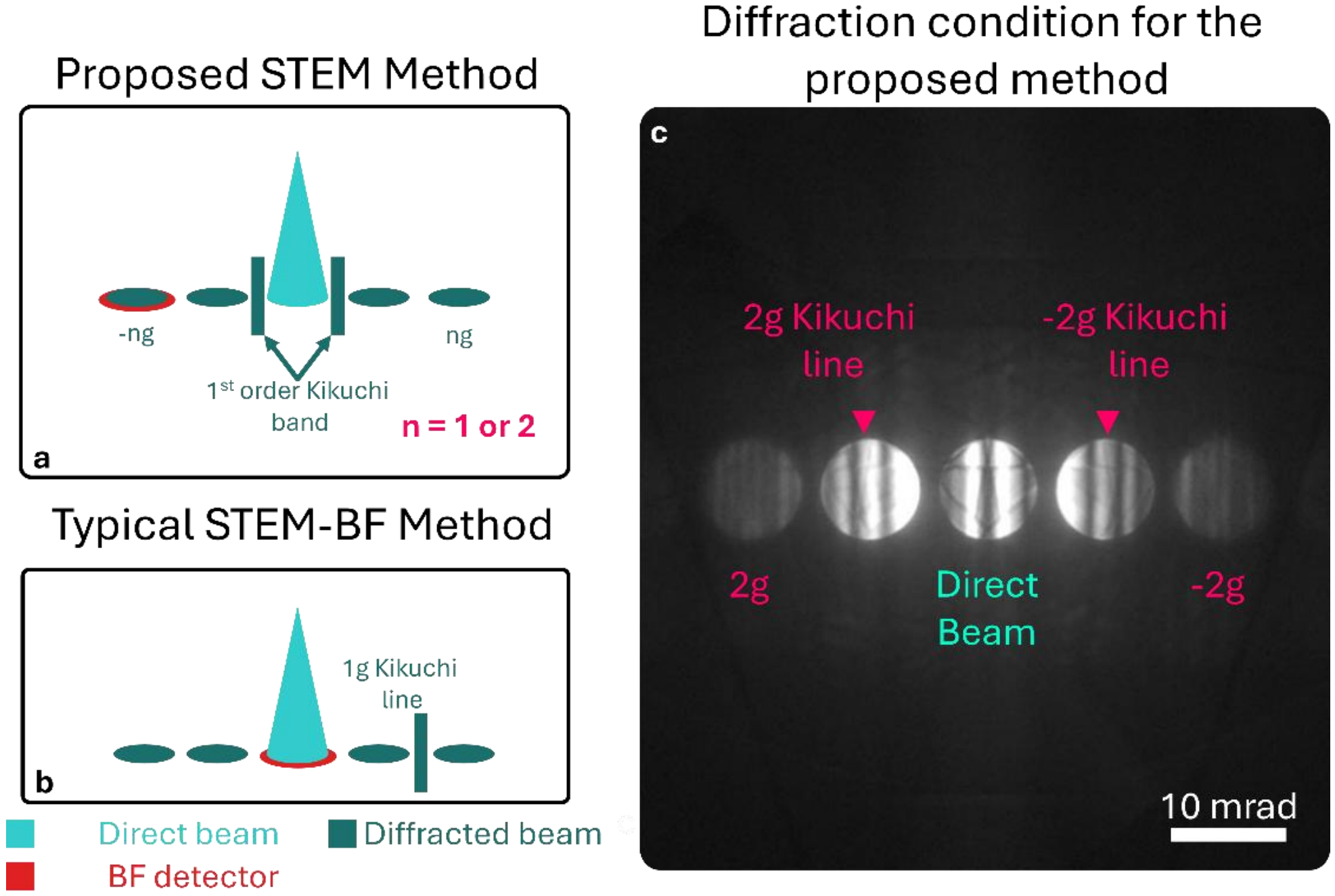

**Fig. 1 Schematic illustration of the optics setup and diffraction conditions for our new dislocation imaging method.** (a) The CBED pattern and optical setup in our proposed method: One micrograph in the region of interest is formed by the -(n+1)**g** disc on the bright-field (BF) detector. (b) Schematical illustration of classical two-beam condition for dislocation imaging in STEM as a comparison. The 1g Kikuchi line is between the 1**g** and 2**g** diffraction spots in the electron diffraction pattern to achieve positive s for g beam and the direct beam is on the BF detector. (c) A corresponding CBED pattern for the proposed method.

# Results

## Demonstration of 0.5-nm resolution, light-element sensitivity superior to atomic STEM-HAADF imaging, and discovery of self-disguised precipitates

Near the 001 zone axis, Fig. 2a-b were recorded using direct disc under s > 0 conditions. Since the pink-circled defects showed residual/invisible and visible contrast (Fig. 2a), which is identical to an edge-on **[100]** dislocation loop exhibiting inside contrast (pure black within a loop), in our initial analysis we assumed these defects to be dislocation loops using classical bright-field (BF) diffraction contrast. Fig. 2c clearly shows parallel fringes with 0.5-nm width within the defect, demonstrating that our new method can provide near atomic level resolution, a new record for diffraction contrast imaging. Because of these periodic fringes, these defects cannot be perfect **[100]** loops in Fe alloys. Consequently, we imaged them with atomic-resolution High-Angle Annular Dark-Field STEM (STEM-HAADF) (fig. 2d), revealing two crystal structures. The matrix is denoted by orange spheres representing a BCC unit cell. The unit cell from the defect is denoted by blue spheres (Fig. 2d). Along **[020]** (white dashed arrow in Fig. 2d), two cells have an identical unit cell parameter (0.28 nm) whereas along **[200]**, the lattice parameter of the defect is 0.42 nm. This indicates that, compared with the matrix, the atomic displacement induced by the defect occurred along **[200]**. Therefore, this defect represents a new phase.

Without our new method, this new phase in the BCC FeCr system would likely not have been discovered. As discussed above, the atomic displacement of the defect is along **[200]** (identical to **[100]**). It is well recognized that pure-edge **[100]**(200) dislocation loops can form in BCC Fe alloys (*13, 14*), which have identical projection with the defect in Fig. 2. Because atomic displacement by this defect is parallel to the displacement by **[100]**(200) loops, visibility/invisibility criteria of the precipitates and **[100]** loops are identical using **g·b** method, which is the traditional basis for dislocation analysis. This provides evidence that the defect only induced displacement along **[200]** but the defect's lattice parameters are identical with the BCC matrix along **[020]** and **[002]**. Thus, we can conclude that the defect has a BCT structure (a=b= 0.28 nm and c=0.42 nm). Classical theories regarding the formation of BCT require austenite (FCC) as the prerequisite for transformation to BCT (*15*). To our knowledge, this is the first observation of BCT formation without stable austenite as prerequisite. Our new method also demonstrated a sensitivity superior to atomic STEM-HAADF equipped with $C_s$ corrector. A precipitate was captured by the new method showing inside fringes (Fig. 2e). To determine its atomic structure, we imaged it with atomic-resolution STEM annular bright field (STEM-ABF) and STEM-HAADF. Unlike the aforementioned BCT precipitate, this precipitate adopted a BCC structure (Fig. 2f-h). This type of irradiation-induced precipitate could not be identified by conventional diffraction contrast, because it exhibited identical visibility/invisibility behavior and shape with irradiation-induced <100> dislocation loops, nor by atomic-resolution STEM-HAADF, because it was fully coherent with the matrix (Fig. 2h). However, STEM-ABF images (Fig. 2f and g) displayed three dark bands along the long side of the precipitate. The difference between atomic HAADF and ABF micrographs indicated the presence of light elements (e.g., C, N, O), subsequently confirmed by electron energy-loss spectroscopy (EELS). The discovery of these two types of precipitates demonstrates that, even without a probe corrector, our method not only achieves 0.5-nm resolution but also detects lattice distortions induced even by light elements.

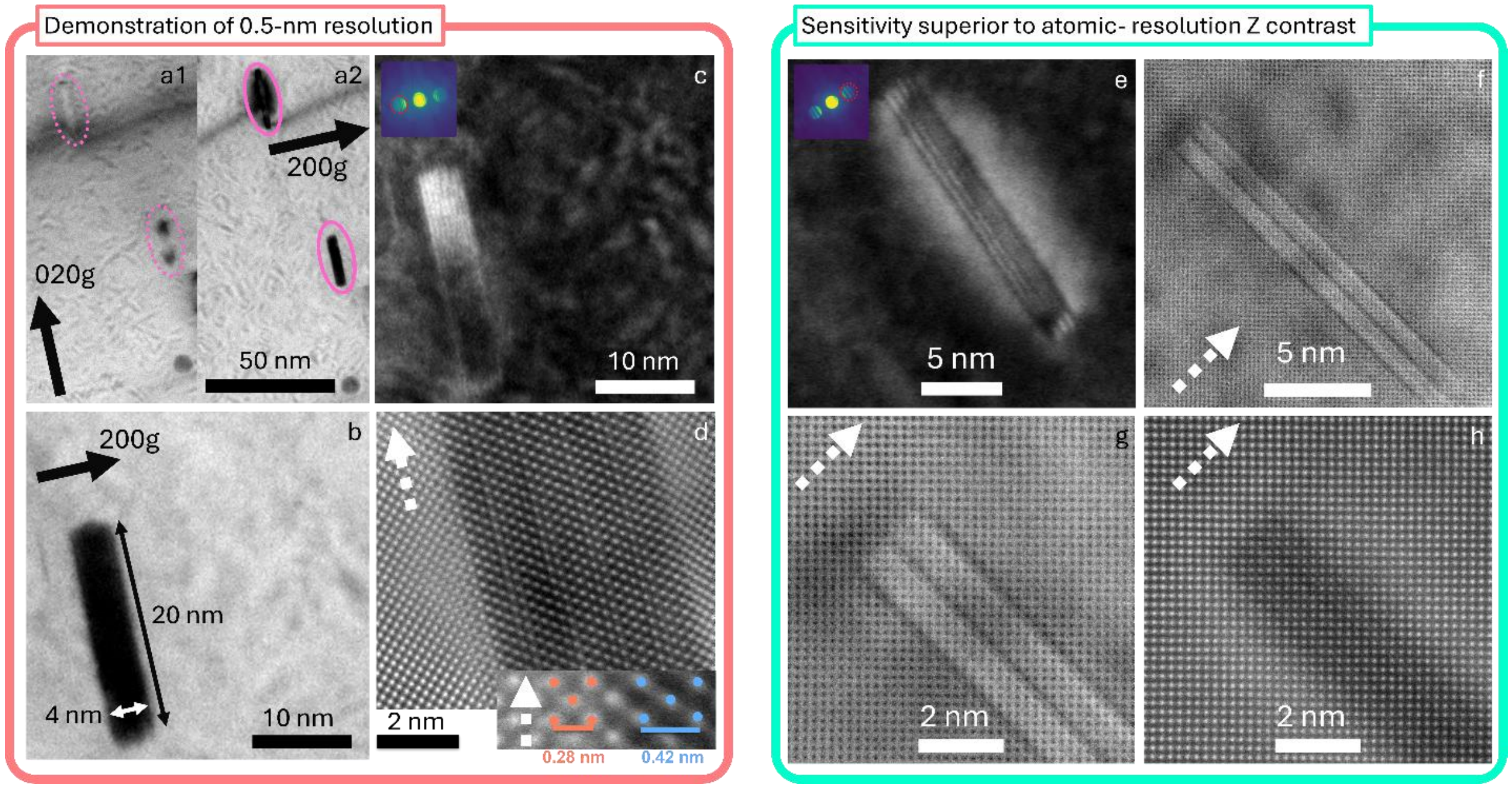

**Fig. 2 Demonstration of high resolution (a-d) and superior sensitivity to atomic displacement (e-h).** (a1-a2) show an irradiation-induced defect (4 × 20 nm). The pink circles highlight two precipitates of interest, which is in residual contrast and invisible condition in (a1) but is visible in (a2). (b) shows a precipitate in higher magnification. (c) With our new method, the defect in (a-b) shows periodic 0.5-nm fringes. (d) The atomic-resolution STEM-HAADF micrograph displays two regions: atoms on a brighter background (matrix) and atoms on a darker background (the defect). The inset in (d) is magnified region from across the matrix-defect interface. (e-h) demonstrate the superior sensitivity compared to atomic-resolution STEM-HAADF. Another precipitate were imaged with our new method (e), showing 0.5-nm fringes. The atomic-resolution STEM-ABF micrographs (f-g) of the precipitate in (e). (f) and (g) are in lower and higher magnifications, respectively. The atomic-resolution STEM-HAADF micrograph (h) of the precipitate in (e). The black and white solid arrows denote two-beam diffraction conditions while the white dashed arrows in (d, f, g, and h) denote 020 crystal direction in BCC system. The red circles in two insets denote the STEM-BF detector.

# Improved reliability of diffraction contrast via paired micrographs

As shown above, a micrograph from single convergent beam electron diffraction (CBED) disc can extract useful data with 0.5-nm resolution. Using paired CBED discs can yield even richer information. Our paired micrograph method involves using opposite CBED discs (e.g. **200g** and $\mathbf{\overline{2}00g}$ or **220g** and $\mathbf{\overline{2}\overline{2}0g}$ in BCC Fe) respectively for micrograph formation. The general idea is similar to conventional inside-outside contrast (IOC), but the critical difference is to keep s equal in our method, e.g., $s_{\mathbf{220g}} = s_{\mathbf{-2-20g}}$. In such case, the user should not tilt the sample but instead shift the CBED pattern and cast the disc of interest on the BF detector. Fig. 3 shows defect images in two paired STEM micrographs from one region. Fig. 3b was recorded after Fig. 3a followed by CBED pattern shift. No tilting was performed. A white arrow highlights a 1.5 nm loop visible as a white dot (inside contrast) in Fig. 3a and outside contrast in Fig. 3b using **110g** systematic row. Its low brightness in both micrographs means it could be mistakenly treated as 'residual contrast' if only one micrograph were available, demonstrating our method's improved accuracy for nano-loops. A cyan arrow marks a pentagonal loop, appearing in Fig. 3a but nearly vanishing in Fig. 3b. Due to its strong variation, it is visible under **110g** row. We will call this 'strong-weak' contrast hereafter. It reveals a novel phenomenon: Loops yielding $\mathbf{g} \cdot \mathbf{b} \neq 0$ appear invisible under a given **g**, but visible under its corresponding negative **g**, which will be addressed below. Both green and pink arrows mark ~2 nm loops that exhibit dissimilar behavior in Fig. 3. Their images indicate that the green-arrowed loop is inclined, but the red-arrowed loop is edge-on. The images of white-, green-, and pink-arrowed loops are consistent with conventional loop image, but the images of the cyan-arrowed loop are not, revealing an unconventional loop.

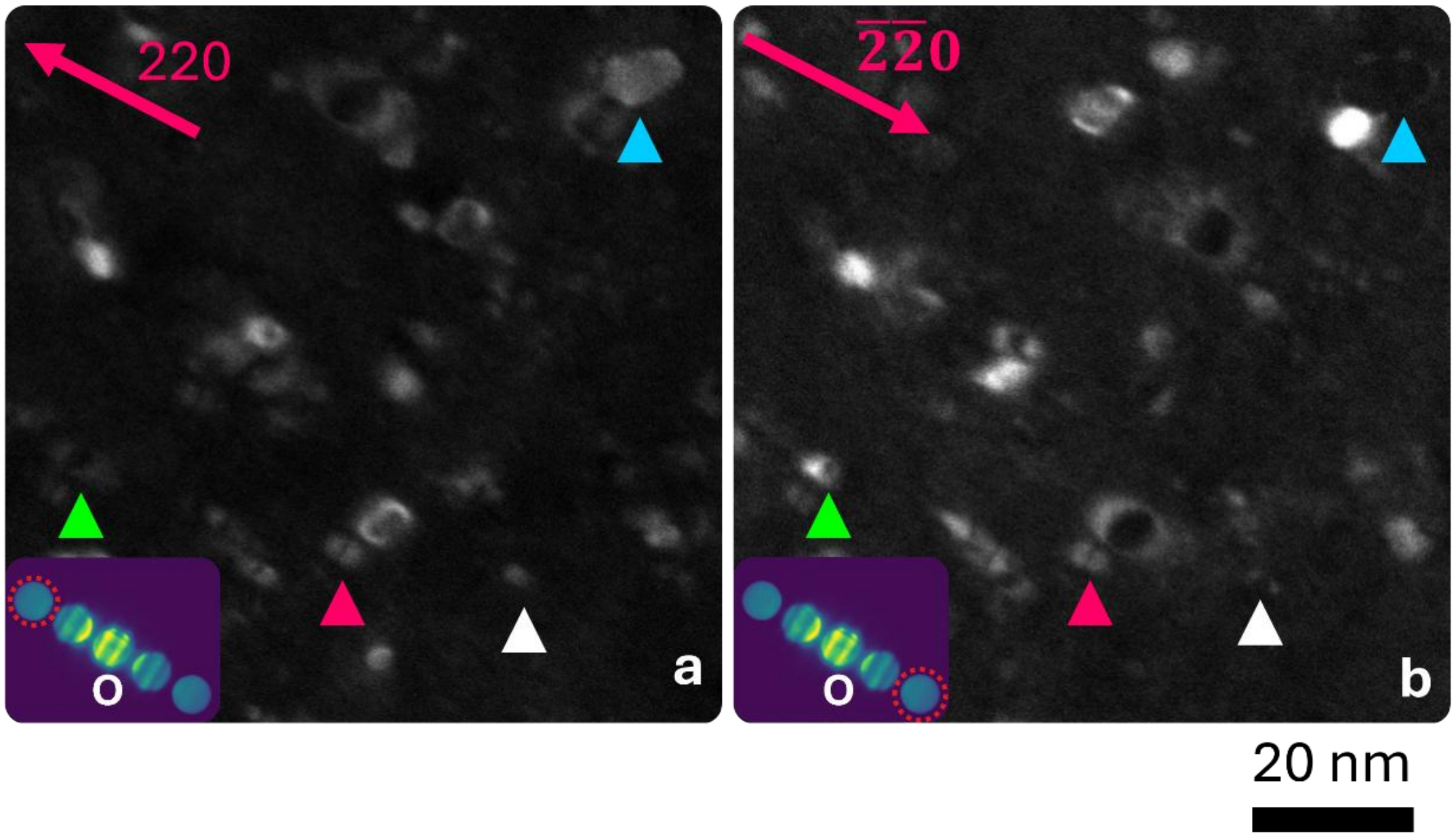

**Fig. 3 A region imaged with two opposite second-order reflection discs.** (a) shows the region formed by **220g** disc on the BF detector and (b) shows the region formed by $\mathbf{\overline{2}\overline{2}0g}$. Four loops are highlighted for illustrative purposes. The loops marked by green and pink arrows are ~ 1-2 nm in diameter. Although they are similar in size, they behave dissimilarly in both micrographs due to their habit planes. The loop denoted by the cyan arrow is pentagonal in (a) but exhibits very weak residual contrast in (b). The loop indicated by the white arrow is ~ 1.5 nm in diameter. By comparing its images in the two micrographs, it is definitely determined to be 'visible,' demonstrating the resolving power of our method. Two insets are the diffraction condition for imaging. Note that the diffraction conditions for (a) and (b) are identical. 'O' refers to the direct disc in the insets. The red-circled disc in each inset is the disc on BF detector for image formation. B ~001Z.

# Discussion

## Contributors of the new-record angstrom resolution

As explained in the above context, the classical (g,3g) WBDF technique achieves ~ 5nm resolution in modern microscopes due to their low Ewald sphere curvature if electrons are at 200-300 kV. Conversely, our proposed diffraction method achieves 0.5-nm resolution under the same voltage. This breakthrough can be primarily attributed to (1) directional focused probe along atomic planes of interest due to our equal-s diffraction condition, and (2) limited interacting volume with defects because of WBDF (Fig. 4). The critical contributor is the fine probe effect. Modern STEMs generate well-focused beams even without $C_s$ corrector. With a semi-convergence angle of ~5.5 mrad, the STEM probe can be ~0.3 nm in 200-kV STEMs if $C_s$ = 1.2 mm (*16, 17*). We simulated the probe in BCC iron (lattice parameter = 0.28 nm) under our method with $C_s$ = 1mm (fig. 4a-c). Probe shape at 8° away from the 001 zone axis (Fig. 4a) and for two other tilts are documented in the Supplementary. Fig. 4 demonstrates that if the equal-s condition ($s_{\mathbf{110g}} = s_{\mathbf{-1\text{-}10g}}$) is satisfied, the simulated in-crystal probe elongates along $\mathbf{[1\overline{1}0]}$ but remains focused along

**[110]** (Fig. 4c), which is the excited systematic row. The probe enters the specimen with an initial full-width half-max (FWHM) of 1.89 Å. The incident wavevector acquires a strong component along **[110]**, leading to preferential coupling into Bloch states associated with these planes. This produces anisotropic channeling, in which the electron intensity is redistributed by crystal potential. The incident probe is decomposed into a superposition of eigenstates set by the tilt condition. The observed real-space redistribution arises from the interference and differential propagation of a limited subset of strongly excited Bloch waves, rather than a lateral focusing action by the lattice. Particularly, modes with transverse character aligned to the (110) planes dominate, giving rise to confinement along **[110]** while remaining comparatively delocalized along the [$\mathbf{1\bar{1}0}$] (Fig. 4e-f). This results in a sub-probe-width localization (~0.5 Å) along **[110]**. This shape remains up to 80-nm thickness, the thickest sample in our simulation. However, the localization invariance with foil thickness occurs at the cost of useable intensity. The usable signal, providing superior resolution, reduces with increasing thickness and is driven primarily by thermal diffuse scattering and the accumulation of incoherent background. The contrast associated with the coherently channeled components diminishes, leading to a monotonic decline in usable signal-to-noise ratio. To generate 0.5-nm resolution at 200kV, we recommend TEM foils < 60 nm in thickness for imaging, whose normalized peak intensity is > 0.5 (fig. 4f).

The next requirement for 0.5-nm resolution is localized diffraction. With $s \approx 0$, the atomic planes of interest are slightly away from Bragg condition so that a small lattice distortion could restore the Bragg condition. This small distortion can be distant from its origin, leading to an image width of up to 20 nm (green region in Fig. 4d). In contrast, for WBDF requiring $s << 0$ or $s >> 0$ (*18*), electrons weakly interact with perfect atomic planes of interest, since the atomic planes are far away from the Bragg condition. To satisfy the Bragg condition, highly distorted regions are needed. Therefore, the diffraction only occurs in the region much closer to its origin (the magenta region near a dislocation core in Fig. 4d). With a well-focused in-crystal probe coupling with WBDF, a defect can only be seen once the probe is exceedingly close to the defect. This inherent advantage of fine probe coupled with WBDF also generates another benefit for microscopists. The precipitate in our FeCr model alloy, shown in Fig. 2e, contained a certain amount of other elements other than Fe and Cr, such as C, N, O on interstitial sites, total concentration >4% (The precipitates in Fig. 2 are irradiation-induced precipitates, and its formation mechanisms will be addressed elsewhere). A new method to image light elements in a small volume is atomic-resolution STEM-ABF (*19*). By such method, we verified that light elements existed in the precipitates, leading to the image of three black bands in Fig. 2f-g. However, this method requires state-of-art instruments equipped with $C_s$ corrector and segmented detectors, and to avoid artificial atoms in atomic micrographs it also requires suitable sample thickness and proper optics setup. Atomic-resolution STEM-HAADF can avoid the artificial atoms in STEM-ABF, but since STEM-HAADF depends on atomic numbers Z, it failed to image these light elements (Fig. 2h). Conversely, our method is diffraction-based, and it elastically probes atomic displacements, which can even be induced by light elements, especially when they are on interstitial sites. Thus, for a nanosized defect containing light elements, our method outperforms STEM-HAADF, requiring intermediate-to-high atomic numbers, and STEM-ABF, depending on phase contrast strictly affected by sample thickness.

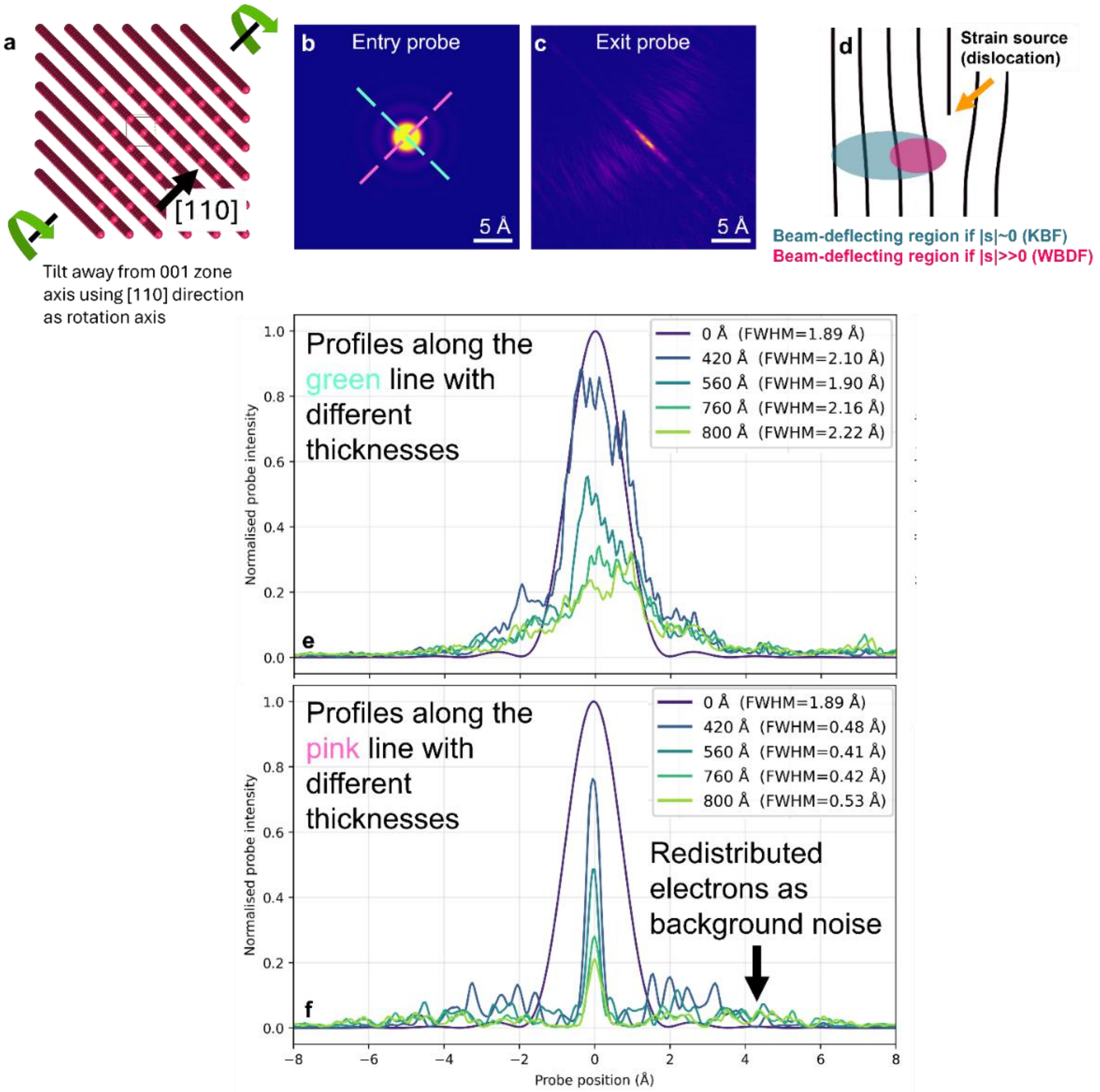

**Fig. 4 Illustration of mechanisms for superior resolution achieved by the new method.** (a) A width comparison of beam-deflecting regions between diffractions with small and large excitation errors. Note that we use the absolute value instead of the value of excitation errors in (a). (b) A comparison of projected widths of distorted atomic planes under (b1) (g,3g) WBDF and (b2) under the new diffraction condition. (c-g) beam shape simulation under our new method. (c) BCC Fe is tilted away from 001 zone axis with 200g systematic row being excited. The black box denotes a BCC unit cell. The inset is the diffraction condition.

In summary, electrons in STEMs for diffraction contrast are highly sensitive to local atomic strains. This high sensitivity is a double-edged sword since strain sources can generate strain fields propagating over distances significantly larger than the sources' physical sizes, causing the imaged size of strain sources to be up to twenty times larger, i.e. poor resolution. Traditional wisdom aimed to weaken electron-solid interactions to restrict the sensitivity only to highly strained regions near the sources: This is the principle behind WBDF. Unfortunately, the classical (g,3g) WBDF technique's resolution with 200-kV TEMs stops at ~ 5nm. An inherent drawback in classical WBDF is that, with increasing s, it leads to a significant beam broadening. Certainly, a method for further resolution improvement is to maintain the highly-strained-region sensitivity but reduce detrimental beam broadening. With this goal in mind, we developed a new widely deployable diffraction condition, i.e. $s_{-g} = s_{+g}$, to focus the electron probe inside crystals along the crystal directions of interest through channeling effects. A new two-beam diffraction contrast record with routine 0.5-nm resolution was achieved and our method shows high sensitivity to light elements like C, N, and O once they induce atomic displacements.

## Enhanced reliability from the new method and advanced imaging on nanosized defects

The superior resolution paves the pathway to address a longstanding question: What is the structure of a defect at stage of nucleation? A good example is irradiation-induced nanosized self interstitial atom (SIA) clusters or equivalently, nanosized loops. Classical diffraction contrast in (S)TEM can detect them without structural information so that they all were regrettably designated as incompletely resolved 'black spots' over the past 50 years. It is possible to prepare ~10-nm thick metallic TEM foils, which are comparable to the nanosized loops, but the success rate for such metallic foils is low so that ~3-nm defects are usually embedded in TEM foils whose thickness can be ten times higher (~ 30-70 nm) than the defect size and the nanosized defects' structures can be 'averaged' by the perfect lattice above and below, leading to failure in projecting the structures in atomic-resolution STEM-HAADF. Using our paired micrograph method, which generated Fig. 3, this longstanding nanodefect characterization challenge can be addressed. This new capability can be qualitatively understood with kinematical two-beam approximation. The diffracted beam of interest $\psi_{ng}$ depends on $\int_0^t \exp[2\pi i(sz - n\boldsymbol{g} \bullet \boldsymbol{R})]\, dz$, where t is the thickness of the TEM foil, s is the excitation error, z is the in-foil defect depth along the probe direction, meaning that z is affected by tilting (*20*), and **R** describes atoms displaced from their ideal positions. We set the TEM foil to the position where the ±**g** Kikuchi lines were *evenly* separated by the direct disc ensuring equal-s condition. Due to large s and STEM mode, our method effectively suppressed the dynamical effects (*21*). Due to the identical value of $s_{+n\mathbf{g}}$ and $s_{-n\mathbf{g}}$ and a procedure free of tilt (no change in z in $\psi_g$'s equation), the sz term in the equation above is identical for $\psi_{+n\boldsymbol{g}}$ and $\psi_{-n\boldsymbol{g}}$. Consequently, the loop imaging size variation from tilting between the paired dislocation loop images was eliminated. Since paired micrographs were captured with ±n**g**, the identical magnitude of +n**g** and -n**g** ensure that the n$\boldsymbol{g} \bullet \boldsymbol{R}$ term above depends on **R** and its corresponding angle from n**g**. In other words, the difference exclusively depends on **R**. As shown in Fig. 5, when imaging a dislocation loop under +n**g** and -n**g**, respectively, the images originate from different sections of the lattice encompassing the dislocation. Therefore, the shape difference between two loop images by respective +n**g** and -n**g** faithfully reflected how loops distort neighboring lattice planes in different fashions, providing novel structural information as highlighted in examples below.

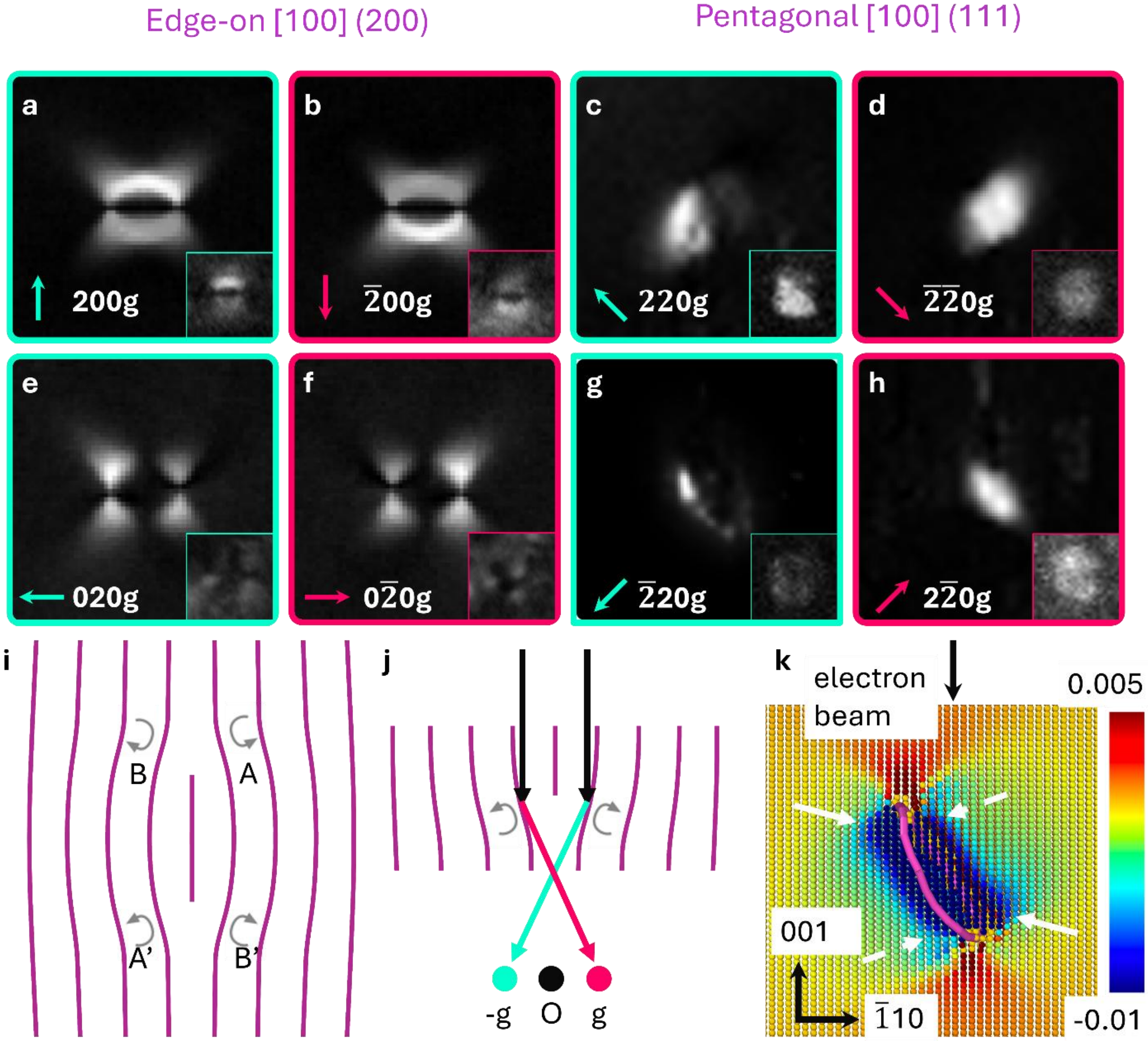

**Fig. 5 Image simulation on selected nano-loops and the origin of abnormal contrast.** Simulated images from an edge-on **[100]**(200) loop are under excited **200g** row (a-b) and **020g** row (e-f). Simulated images from a **[100]**(111) loop are under excited **110g** row (c-d) and **1$\bar{1}$0g** row (g-h). The arrow and number in each simulated image represent the diffracted beam for image formation. Each inset displays the corresponding experimental result. Loops strongly appear in (a-d) and h but nearly vanish in (e-g). (i) A pure-edge loop induces two sets of (counter-) clockwise-bent atomic planes, labelled as B (A) and B' (A'). (j) displays the lower dislocation core in (i) and its relationship between atomic planes' bending direction, indicated by arrows, and electron diffractions. (k) A non-edge loop and its strain field. Strains around clockwise-bent atomic (110) planes are highlighted by solid arrows and around counter-clockwise-bent atomic (110) planes are highlighted by dashed arrows. Penta-loop's atomic structure is in the Supplementary file. (j**)** is digitally reproduced from Williams and Carter (*6*).

Well-documented loop shapes are 'coffee bean' contrast from edge-on loops and IOC from inclined loops. These classical images have been obtained from large loops, and all nano-loops are typically categorized as "black spots", without clarification of their structures or imaged shapes. Shown in Fig. 3, our new method reveals that the two classical images are resolvable for several loops down to ~ 1 nm (~ 37 SIAs in BCC iron if b= ½**<111>**), demonstrating that the structures of large loops can also exist at small sizes. However, numerous nano-loops were in several other modes, revealing nano-loops with additional unconventional structures. Fig. 3 records three different loops after **g·b** analysis: edge-on loops along the 001 zone in 'coffee bean' shape, representing edge-type [100] loop; inclined circular loops, representing edge-type ½[111] loops; and a 15-nm pentagonal [100] loop (highlighted by cyan arrows), which has not been previously reported. We performed image simulation on these dislocation loops but with smaller sizes (d = 4nm). The full results are in Fig. S4 with key results highlighted in Fig. 5, with the corresponding insets displaying our experimental observations. Fig. 5a-b show that differences in edge-on **[100]**(200) loops under **200g** and $\bar{2}$**00g** persist and the loop images are solid and continuous, consistent with our experimental observations. (Note that we did not use second order reflections in **200g** row; see Supplementary). For loops, the n**g·R** term in the equation above can be approximated as n**g·b**, where b is the Burgers vector of a loop. If n**g·b** = 0, the loop should be invisible or show residual contrast. Under **g•b** = 0 with **020g** row, residual contrast (Fig. 5e-f) appears discontinuous. Fig. 5c-d show that a pentagonal **[100]**(111) loop (penta-loop for short hereafter) actually shows 'inside-outside' contrast under **220g** and $\bar{2}\bar{2}$**0g** in simulation, but its right section under **220g** is too weak to be distinguishable from the background in simulation and experiment, leading to 'dot-dot' contrast with shape variation. It is noteworthy that the 'dot-dot' contrast is not due to the tiny sizes. As shown in Fig. 3, 'coffee bean' and IOC remain visible on loops with d~1 nm. Another novel observation is that the penta-loop is similar to residual contrast under **g•b** ≠ 0 (Fig. 5g) but visible under **-g•b** ≠ 0 (Fig. 5h). It is also true for the cyan-arrowed loop in Fig.3. This arises from its asymmetrical strains. Fig. 5j shows how electrons are diffracted by the atomic planes around a dislocation core. The counter-clockwise-bent planes on the left of the dislocation reflect electrons into +**g** but the clockwise-bent planes on the right reflect electrons into -**g**. A loop distorts two sites clockwise (B and B') and another two counter-clockwise (A and A') (Fig. 5i). The detected +**g**(-**g**) disc reflects distortion from sites A (B) and A' (B'). For a pure-edge loop, the magnitudes of clockwise and counter-clockwise distortions are similar, yielding symmetrical brightness under ±**g** (Fig. 5a vs 5b). However, as shown in Fig. 5k, if the penta-loop's strains are measured using (110) planes, the clockwise atomic planes (highlighted by the solid arrows) bend more than its counter-clockwise distortion (highlighted by dashed arrows), creating unequal magnitude in diffraction contrast. The difference between Fig. 5g and 5h faithfully reflects asymmetrical distortion by the penta-loop.

In summary, a persistent lack of tools for direct nanoscale two-beam diffraction contrast microscopy characterization has left both spatially close (separated by several nanometers) and nanosized defects largely unexplored, not only impeding our understanding of defect physics but also deterring advanced material design. To address these challenges and advance studies of defective crystalline material, we have demonstrated a new method using STEM-WBDF with equal negative excitation error, providing novel insight into the structure of nano-sized defects or defects' atomic-scale internal arrangements. This new method not only resolved internal arrangements of BCT precipitates with 0.5-nm resolution but also quantified the long-mysterious irradiation-induced 'black spots' down to ~1 nm in BCC FeCr, creating an opportunity to directly compare atomic simulation with experiments with dependable statistics. Under low-to-intermediate magnifications covering dozens of defects in one micrograph, resolving subtle variation in atomic arrangements remains. Microscopists can identify the representative defects among the dozens of ones, serving as a prescreening tool for time-consuming strain mapping by 4D-STEM, which is introduced in the Supplementary file.

# Method

## Sample preparation and dislocation loop production

To introduce dislocation loops, a Fe-8Cr (wt%) sample with very low impurity content was ion irradiated (*22*). Before the irradiation, the Fe-8Cr cube (~2.5-3.5 mm in dimension) had been cut from a bulk rod (~3cm in diameter) by electrical discharge machining (EDM) employing copper wire. The Fe-8Cr cube had been mechanically reshaped as ~2 mm(W) × ~2 mm(L) × ~0.5 mm (H) specimens. The mirror surface preparation for irradiation had involved mechanical grinding utilizing sandpaper up to 2000 grit, finished by 0.05 μm colloidal silica on an Allied Multiprep® polisher. The sample with mirror-like surface was irradiated at The Michigan Ion Beam Laboratory (MIBL). The irradiation was performed by 0.8-MeV protons at 250 °C to a midrange dose of 2 dpa and 8 MeV Fe ion at 450 °C to a midrange dose of 0.35 dpa with wobbling defocused mode. The damage profiles were evaluated using the software, Stopping and Range of Ions in Matter (SRIM) with quick calculation mode and assuming 40 eV displacement energy. The dose rate was $10^{-5}$ dpa/s for both irradiations.

## TEM sample preparation and STEM characterization

To lift out a TEM foil with [001] orientation from the post-irradiation bulk specimen, we applied electron backscatter diffraction (EBSD) technique using by EDAX® Velocity Plus EBSD on a ThermoFisher® Helios 5 SEM. The TEM foil was prepared by focused ion beam (FIB) technique using standard techniques and intentionally lifted out several tens of micrometers away from grain boundaries to minimize grain boundary's impact. The FIB was ThermoFisher® Helios 5 Hydra Dual Beam plasma FIB (PFIB) SEM. Pt cap was deposited 12 kV 1nA Xe ion probe. Liftout started with 30-kV 65-nA Xe ions. The finishing process was performed with a 30-kV 30-pA Xe ions, The final processing was by flash electropolishing method on the TEM foil described in (*23*).

The TEM foil was examined in ThermoFisher® Spectra 300 TEM located at the Institute for Advanced Materials & Manufacturing (IAMM) at University of Tennessee, Knoxville. We operated at 200 kV. 20 µm C2 aperture was inserted and the STEM semi-convergence angle, α, was 5-5.5 mrad. Dependent on the **g** vector for imaging formation, appropriate large camera lengths and objective apertures were applied to ensure that each detector received from one diffracted disc (*24*). Seven types of dislocation loops in two families exist in BCC iron system (*25-29*) and they can be distinguished by the family of **110g** and the family of **200g**. Thus, we tested three different two-beam imaging conditions (**110g**, **$1\bar{1}0$g**, and **200g**, respectively), the conditions that are enough to distinguish ½**<111>** type and **<100>** type. Besides two-beam conditions, on-zone STEM under [001] zone axis micrographs were recorded. Under suitable setup, on-zone STEM can simultaneously record all seven types of dislocation loops in BCC iron system (*30-32*). Table 2 lists the diffraction conditions in this study.

*Table 1 The invisibility criterion in the classical **g•b** method on seven different loop types in BCC iron*

| | 011Z | 001Z |
|---|---|---|

| b \ g | $2\bar{1}1$g | $0\bar{1}1$g | 110g | $1\bar{1}0$g | 200g |
|---|---|---|---|---|---|
| [111] | 2(4, 6) | 0(0, 0) | 2(4, 6) | 0(0, 0) | 2(4, 6) |
| [$11\bar{1}$] | 0(0, 0) | -2(-4, -6) | 2(4, 6) | 0(0, 0) | 2(4, 6) |
| [$1\bar{1}1$] | 4(8, 12) | 2(4, 6) | 0(0, 0) | 2(4, 6) | 2(4, 6) |
| [$\bar{1}11$] | -2(-4, -6) | 0(0, 0) | 0(0, 0) | -2(-4, -6) | -2(-4, -6) |
| [100] | 2(4, 6) | 0(0, 0) | 1(2, 3) | 1(2, 3) | 2(4, 6) |
| [010] | -1(-2, -3) | -1(-2, -3) | 1(2, 3) | -1(-2, -3) | 0(0, 0) |
| [001] | 1(2, 3) | 1(2, 3) | 0(0, 0) | 0(0, 0) | 0(0, 0) |

Numbers in black, bule (the first number in each parentheses), and red (the second number in each parentheses) denote the values of **g•b**, 2**g•b**, and 3**g•b**, respectively. For numbers in each cell, they become larger if a higher order **g** vector is applied in imaging.

The diffraction conditions of our two proposed methods are schematically illustrated in Fig. 3. For method #1, the (n+1)g's Kikuchi line was between the n**g** and (n+1)**g** diffraction discs. In this setup, the excitation error of n**g** disc, $s_{ng} > 0$, but the excitation error of (n+1)**g** disc, $s_{(n+1)g} < 0$. By shifting the CBED pattern and utilizing a suitable objective aperture, the (n+1)**g** disc was on the BF detector and the ng disc was on ADF detector, depicted in Fig. 3a. For method #2, the n**g** Kikuchi line was in the very center position between direct disc and n**g** disc. It simultaneously holds that the -n**g** Kikuchi line was in the very center position between the direct disc and -n**g** disc. After tilting the TEM foil to the diffraction condition, Step two of method #2 is to shift the CBED pattern and cast the -n**g** disc on the BF detector, shown in Fig. 3b. After micrograph acquisition in the regions of interests, shift the CBED again and cast the n**g** disc on BF detector (Step three of method #2), followed by micrograph acquisition in the identical regions, shown in Fig. 3c. Note that no sample tilt occurred between Step two and three. To correlate the inside (outside) contrast with positive (negative) value of (**g•b**)s, we will follow the terminology by Föll (*33*) in the following context, i.e. the outside (inside) contrast should be yielded if (**g•b**)s smaller than 0 (greater than 0) .

*Table 2 A comparison between the classical and the proposed method*

| | CTEM | Proposed STEM method |
|---|---|---|
| Step 1 | Tilt the specimen from on-zone to a systematic row diffraction condition by ~ 8° | |
| Step 2 | Tilt the specimen from systematic row diffraction to a two-beam condition | Adjust the systematic row diffraction condition to ensure that the (n+1)**g** and –(n+1)**g** Kikuchi lines are evenly separated by the direct disc |
| Step 3 | Insert an objective aperture to cover the direct beam and switch to 'image' mode | Adjust camera length and shift the CBED pattern to ensure (n+1)**g** disc on STEM-BF detector and record a micrograph§ |
| Step 4 | Record the image | Shift the CBED pattern to ensure -(n+1)**g** disc on STEM-BF detector and record another micrograph§ |

§: to protect an ADF detector from potential beam damage due to high current in a direct beam, we recommend retracting the ADF detector or blocking the direct beam by an objective aperture in our method.

## Probe shape simulation

Probe broadening simulations were performed using a multi-slice approach implemented in the abTEM library in Python (*34*). The scattering potential is calculated from atomic coordinates of bulk BCC iron with a ~ 3 x 3 x 80 nm simulation cell tilted 8 degrees in the [110] direction from [001] zone axis to match the experimental conditions. The effect of thermal vibrations is approximated using frozen phonons and the final scattering potential was calculated using 1 Angstrom sampling. The incident electron probe was calculated to also match experimental parameters, with semi-convergence angle of 5.5 mrad at 200 keV, and aberration coefficients matching measured experimental aberration coefficients.

## Tiny loop image simulation

We employed an atomic-resolution image simulation approach using the abTEM library in Python for improved accuracy. Atomic coordinates were initialized from a Crystallographic Information Files (CIF) of **[100]**(200), **[100]**(111), ½**[111]**(111) dislocation loop (d = 4 nm) centered in ~ 20 x 20 x 20 nm simulation cell of bulk BCC Fe, W, and V. The square-shaped **[100]**(200) dislocation loop was constructed by inserting two layers of self-interstitial atoms into a perfect bcc lattice. Similarly, the **[100]**(111) loop was created by inserting two layers of self-interstitial atoms into a perfect BCC iron system. The Burgers vectors are along **[100]** using LAMMPS [1, 2]. The interatomic potential developed by Malerba et al. was employed for the loop construction in pure Fe due to its proven performance in radiation damage studies (*35*). The interatomic potentials developed by Mason et al. (*36*) (W) and Chen et al. (*37*) (V) were used, respectively. All structures were first energy-minimized, then equilibrated in the NPH ensemble at 300 K for 1 ns to stabilize the system density and finally relaxed again to their local energy minimum.

From image simulation, the beam is tilted 8 degrees in the [110] direction from [001] zone axis in order to satisfy two-beam approximation. Thermal vibrations were incorporated using the frozen phonon method and the electrostatic potential was calculated using a sampling interval of 0.5 Angstroms. An electron probe with convergence angle 5 mrad and energy 200 keV was defined with slight defocus and spherical aberration to better approximate microscope conditions. The scattering process was modeled using a multi-slice approach, specifically using a scattering matrix with an angular interpolation factor of 4. Sampling was determined based on Nyquist's theorem. The intensity of the scattered electrons was recorded using a simulated pixelated detector with a maximum scattering angle of 60 mrad. Image reconstruction was accomplished by summing the intensity of the -n**g** disc and n**g** disc for each probe position, producing two images representing the relative intensities of each diffracted disc.

# Acknowledgement:

The authors would like to deeply thank Dr. Neal Evans (UTK) for the help with instruments. This work is supported as part of FUTURE (Fundamental Understanding of Transport Under Reactor Extremes) with Los Alamos National Laboratory (BPU), an Energy Frontier Research Center funded by the U.S. Department of Energy (DOE), Office of Science, Basic Energy Sciences (BES). Los Alamos National Laboratory is operated by Triad National Security, LLC, for the National Nuclear Security Administration of U.S. Department of Energy (Contract No. 89233218CNA000001).

This work is also equally supported by DOE, Office of Science, Office of Fusion Energy Science (FES) from grant # DE-SC0023293 with the University of Tennessee (SJZ).

# CRediT authorship contribution statement

Conceptualization: YL, YZ

Investigation: ACH, YL, ZY, ZQ, SC, SJK

Methodology: ACH, YL

Visualization: ACH, YL

Funding acquisition: BQU, SJZ

Supervision: BQU, GD, SJZ

Writing – original draft: ACH, YL

Writing – review & editing: ACH, BKD, BPU, GD, SC, SJK, SJZ, YL, YZ, ZY, ZQ

# Declaration of Competing Interest

The authors declare that they have no known competing financial interests or personal relationships that could have appeared to influence the work reported in this paper.

# Data availability

Data will be made available on request.

# References:

1. C. Liu *et al.*, Two-dimensional superconductivity and anisotropic transport at KTaO3 (111) interfaces. *Science* **371**, 716-721 (2021).
2. R. Lv *et al.*, A 3-GPa ductile martensitic alloy enabled by interface complexes and dislocations. *Nature Materials*, (2026).
3. B. C. Hornbuckle *et al.*, A high-temperature nanostructured Cu-Ta-Li alloy with complexion-stabilized precipitates. *Science* **387**, 1413-1417 (2025).
4. B. B. Zhang, Y. G. Tang, Q. S. Mei, X. Y. Li, K. Lu, Inhibiting creep in nanograined alloys with stable grain boundary networks. *Science* **378**, 659-663 (2022).
5. Z. Chen *et al.*, Electron ptychography achieves atomic-resolution limits set by lattice vibrations. *Science* **372**, 826-831 (2021).
6. D. B. Williams, C. B. Carter, *Transmission Electron Microscopy: A Textbook for Materials Science*. (Springer New York, NY, ed. 2, 2009), pp. 775.
7. D. J. H. Cockayne, Weak-Beam Electron Microscopy. *Annual Review of Materials Science* **11**, 75-95 (1981).
8. Y.-R. Lin, Y. Li, S. J. Zinkle, J. D. Arregui-Mena, M. G. Burke, Application of Weak-Beam Dark-Field STEM for Dislocation Loop Analysis. *Microscopy and Microanalysis* **30**, 681-691 (2024).
9. Q. Pan, L. Lu, Fatigue in metals and alloys. *Nature Materials*, (2025).
10. Q. Pan *et al.*, Gradient cell–structured high-entropy alloy with exceptional strength and ductility. *Science* **374**, 984-989 (2021).
11. D. An *et al.*, The Role of Dislocation Type in the Thermal Stability of Cellular Structures in Additively Manufactured Austenitic Stainless Steel. *Advanced Science* **11**, 2402962 (2024).
12. B. Guo *et al.*, Segregation-dislocation self-organized structures ductilize a work-hardened medium entropy alloy. *Nature Communications* **16**, 1475 (2025).
13. Y. Li, Z. Qi, A. Bhattacharya, S. J. Zinkle, Temperature and dose effects on dislocation loops in self-ion irradiated high-purity iron. *Acta Materialia* **296**, 121235 (2025).
14. R. Schäublin, B. Décamps, A. Prokhodtseva, J. F. Löffler, On the origin of primary ½ $a_0$ <111> and $a_0$ <100> loops in irradiated Fe(Cr) alloys. *Acta Materialia* **133**, 427-439 (2017).
15. Y. Li, D. S. Martín, J. Wang, C. Wang, W. Xu, A review of the thermal stability of metastable austenite in steels: Martensite formation. *Journal of Materials Science & Technology* **91**, 200-214 (2021).
16. M. Weyland, D. A. Muller, Tuning the convergence angle for optimum STEM performance. *arXiv preprint arXiv:2008.12870*, (2020).
17. R. F. Egerton, M. Watanabe, Spatial resolution in transmission electron microscopy. *Micron* **160**, 103304 (2022).
18. D. J. H. Cockayne, A Theoretical Analysis of the Weak-beam Method of Electron Microscopy. *Zeitschrift für Naturforschung A* **27**, 452-460 (1972).
19. K. Ooe, T. Seki, Y. Ikuhara, N. Shibata, High contrast STEM imaging for light elements by an annular segmented detector. *Ultramicroscopy* **202**, 148-155 (2019).
20. B. Fultz, J. M. Howe, *Transmission Electron Microscopy and Diffractometry of Materials*. (Springer Berlin Heidelberg, Berlin, Heidelberg, ed. 3rd, 2008), pp. 758.
21. D. J. H. Cockayne, The principles and practice of the weak-beam method of electron microscopy. *Journal of Microscopy* **98**, 116-134 (1973).

22. Y. Li, G. D. Parker, Y. Zhao, X.-Y. Yu, S. J. Zinkle, Temperature and dose rate effects on dislocation loop formation and decoration in self-ion irradiated high-purity Fe-Cr model alloys. *Acta Materialia*, 122435 (2026).
23. Y. Li *et al.*, Flash electropolishing of BCC Fe and Fe-based alloys. *Journal of Nuclear Materials* **586**, 154672 (2023).
24. C. J. Humphreys, Fundamental concepts of stem imaging. *Ultramicroscopy* **7**, 7-12 (1981).
25. B. C. Masters, Dislocation Loops in Irradiated Iron. *Nature* **200**, 254-254 (1963).
26. S. J. Zinkle, B. N. Singh, Microstructure of neutron-irradiated iron before and after tensile deformation. *Journal of Nuclear Materials* **351**, 269-284 (2006).
27. E. A. Little, B. L. Eyre, The geometry of dislocation loops generated in α-iron by 1 MeV electron irradiation at 550°C. *Journal of Microscopy* **97**, 107-111 (1973).
28. L. L. Horton, J. Bentley, J. Farrell, A TEM study of neutron-irradiated iron. *Journal of Nuclear Materials* **108&109**, 222-233 (1982).
29. S. I. Porollo, A. M. Dvoriashin, A. N. Vorobyev, Y. V. Konobeev, The microstructure and tensile properties of Fe–Cr alloys after neutron irradiation at 400°C to 5.5–7.1 dpa. *Journal of Nuclear Materials* **256**, 247–253 (1998).
30. C. M. Parish, K. G. Field, A. G. Certain, J. P. Wharry, Application of STEM characterization for investigating radiation effects in BCC Fe-based alloys. *Journal of Materials Research* **30**, 1275-1289 (2015).
31. P. J. Phillips, M. C. Brandes, M. J. Mills, M. De Graef, Diffraction contrast STEM of dislocations: Imaging and simulations. *Ultramicroscopy* **111**, 1483-1487 (2011).
32. P. J. Phillips, M. J. Mills, M. De Graef, Systematic row and zone axis STEM defect image simulations. *Philosophical Magazine* **91**, 2081-2101 (2011).
33. H. Föll, M. Wilkens, A simple method for the analysis of dislocation loops by means of the inside-outside contrast on transmission electron micrographs. *physica status solidi (a)* **31**, 519-524 (1975).
34. J. Madsen, T. Susi, The abTEM code: transmission electron microscopy from first principles. *Open Research Europe* **1**, 13015 (2021).
35. L. Malerba *et al.*, Comparison of empirical interatomic potentials for iron applied to radiation damage studies. *Journal of Nuclear Materials* **406**, 19-38 (2010).
36. D. R. Mason, D. Nguyen-Manh, C. S. Becquart, An empirical potential for simulating vacancy clusters in tungsten. *Journal of Physics: Condensed Matter* **29**, 505501 (2017).
37. Y. Chen *et al.*, Interatomic potentials of W–V and W–Mo binary systems for point defects studies. *Journal of Nuclear Materials* **531**, 152020 (2020).
38. H. Kohl, L. Reimer, *Transmission Electron Microscopy*. (ed. 5th edition, 2008).
39. Y. Li *et al.*, Nanoscale Origins of Mesoscale Organization: Creation, Evolution, and Self-Healing of Self-Assembled Dislocation Structures. *SSRN*, (2026).

# Supplementary file

## The validation of inside-outside contrast with negative excitation errors in STEM and a method for large loops (diameter > ~8 nm)

We confirmed that the inside-outside contrast is consistent in CTEM and STEM (*8*). Since it is unusual to form diffraction contrast with s<0, the first task was to validate our method's consistency with the conventional inside-outside approach. A comparison between the conventional and our approach is in Table S1. To achieve s<0, we set the diffraction condition as shown in Fig. S1. Fig. S2a-S2d display the inside-outside contrasts of a 'fish'-shaped ½<111> loop under conventional conditions. We tilt the sample to two different two-beam conditions: excited $\mathbf{1\bar{1}0g}$ (s>0) and $\mathbf{\bar{1}10g}$ (s>0), respectively. Two sets of STEM micrographs (Fig. S2a-S2b and Fig. S2c-S2d) are recorded. The loop exhibits an outside contrast under $\mathbf{1\bar{1}0g}$ and inside contrast under $\mathbf{\bar{1}10g}$. Next, we tilt the sample to the diffraction condition, under which the $\mathbf{3\bar{3}0}$ Kikuchi line was tangent with $\mathbf{3\bar{3}0g}$ disc. Since the high order disc was weak, the signal from thermally diffused electrons could potentially downgrade its micrograph quality. To avoid this, the camera length was set at 720 mm and CBED was shifted so that BF detector exclusively received the signal from the **3g** disc, functioning as detector and 'aperture' simultaneously, and the ADF detector mainly received the **2g** disc with the insertion of a proper objective aperture. The ½<111> loop appeared outside contrast with $\mathbf{2\bar{2}0g}$ disc (Fig. S2e), consistent with the result in Fig S2a-S2b since both yielded identical sign of (**g•b**)s. In contrast, the ½<111> loop appeared inside contrast within $\mathbf{3\bar{3}0g}$ disc due to the $s_{3\bar{3}0}$< 0 (Fig. S2f). Without tilting, we not only achieved the inside and outside contrast from one loop simultaneously, but also our results were consistent with the results by conventional imaging conditions ($s_g$ > 0). Fig. S2e-S2h demonstrate the superiority of our method. The **g·b** value table in Method section exhibits that no ½<111> loops can satisfy **g·b** ≠ 0 with **110g** and $\mathbf{1\bar{1}0g}$ rows. The ½ <111> loop remains visible in all micrographs, proving the relaxation of conventional invisibility criterion. But with $\mathbf{1\bar{1}0g}$ row excited, outside contrast forms in the 2**g** ($\mathbf{2\bar{2}0g}$) image with $s_{2g}$>0 (Fig. S2e) and inside contrast forms in the 3**g** ($\mathbf{3\bar{3}0g}$) image with $s_{3g}$<0 (Fig. S2f). With **110g** row, no significant difference in loop's image is in the 2**g** micrograph ($s_{2g}$>0) and the 3**g** micrograph ($s_{3g}$<0). By comparing Fig. S2e- S2h, it is evident that the 'fish' loop yields **g·b** = 0 if **110g** is excited.

In our methods, we replace the confusing invisibility criterion by outside vs inside contrast. In our method #1, we imaged loops using two unilateral **g** vectors, such as **220g** and **330g**. The signs of **220•b** and **330•b** were identical but opposing signs of (**g•b**)s in the two simultaneously acquired micrographs due to opposite signs of $s_{\mathbf{220g}}$ and $s_{\mathbf{330g}}$. To achieve this goal, the TEM foil was under diffraction condition illustrated in Fig. 6d and we generated the outside and inside contrast of a given loop simultaneously in STEM (Fig. 3). In contrast, the results would change for loops yielding **g•b** = 0. Whatever the s value is, the image of a **g•b** = 0 loop would be on its core and the size variation between its images in the **220g** micrograph and **330g** micrograph would be negligible or identically zero. However, method #1 could be of limited utility due to the size variation arising from the order of **g** discs. The image width of a dislocation is proportional to $\frac{\xi_{eff}}{3\pi s_g}$ (*6*), where $\xi_{eff}$ is the effective distinction distance and $s_g$ is the excitation error of the diffracted disc of interest. An increase in the order of **g** vector in a systematic row leads to an increase in its extinction distance. The change from n**g** to (n+1)**g** leads to an increase in dislocation image width. Due to this change, the size variation could be comparable to the image size of tiny loops, undermining the accuracy of method #1 on tiny loops.

*Table S1 Operations of the classical method and the proposed method for large loops*

| | CTEM | STEM method 1 for large loops illustrated in Fig.S1 | STEM method 2 in main text illustrated in Fig.1 |
|---|---|---|---|
| Step 1 | Tilt the specimen from on-zone to a systematic row diffraction condition | | |
| Step 2 | Tilt the specimen from systematic row diffraction to a two-beam condition | Tilt the specimen from systematic row diffraction to the two-beam kinematical condition that the (n+1)**g** Kikuchi line tangent to (n+1)**g** disc | Adjust the systematic row diffraction condition to ensure that the (n+1)**g** and –(n+1)**g** Kikuchi lines are evenly separated by the direct disc |
| Step 3 | Insert an objective aperture to cover the direct beam and switch to 'image' mode | Insert an objective aperture to cover n**g** and (n+1)**g** discs and adjust camera length to ensure that STEM-BF and -ADF detector only receive one diffracted disc, respectively | Adjust camera length and shift the CBED pattern to ensure that (n+1)**g** disc on STEM-BF detector and record a micrograph |
| Step 4 | Record the image | Record two images by both detectors | Shift the CBED pattern to ensure that -(n+1)**g** disc on STEM-BF detector and record another micrograph |

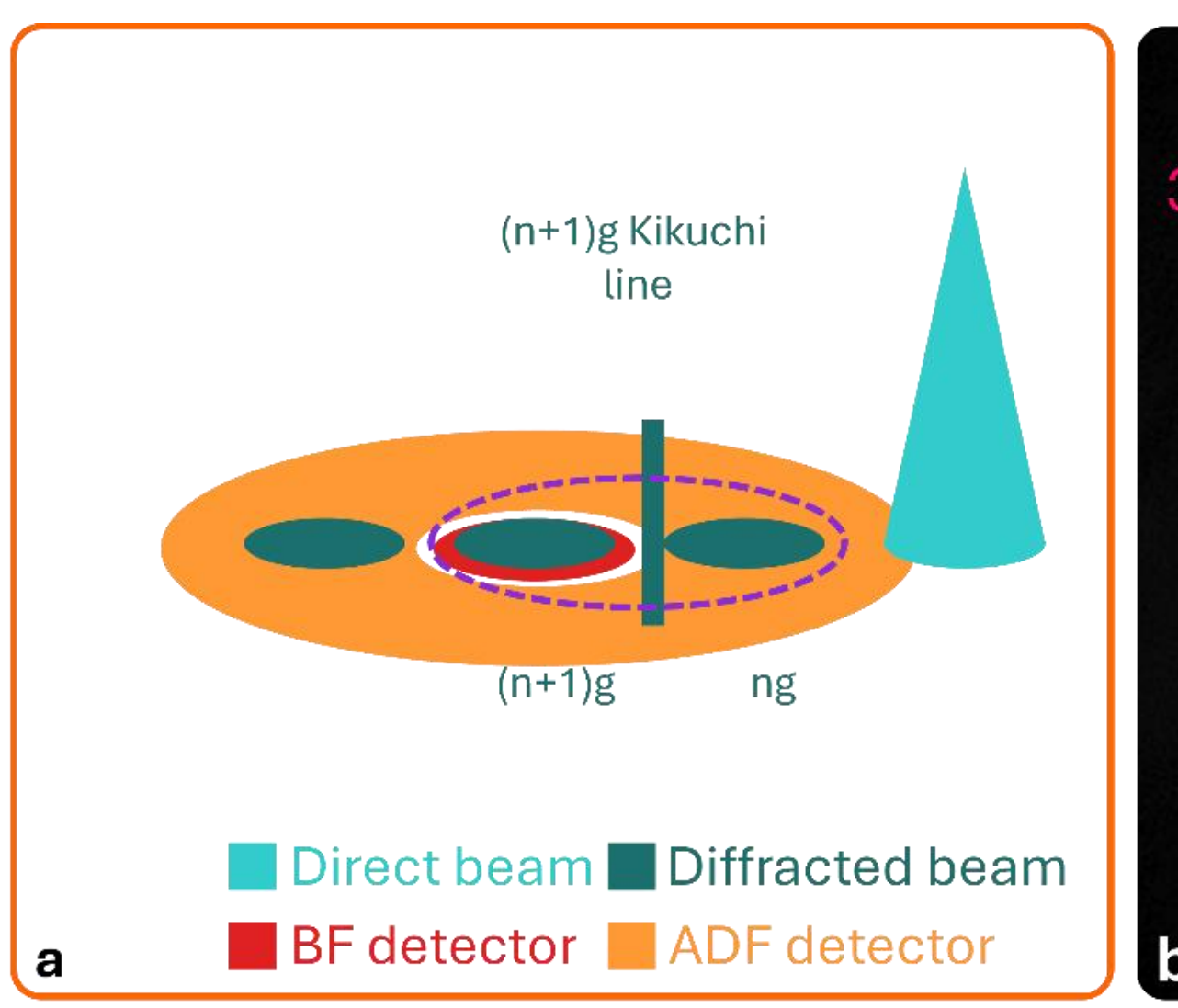

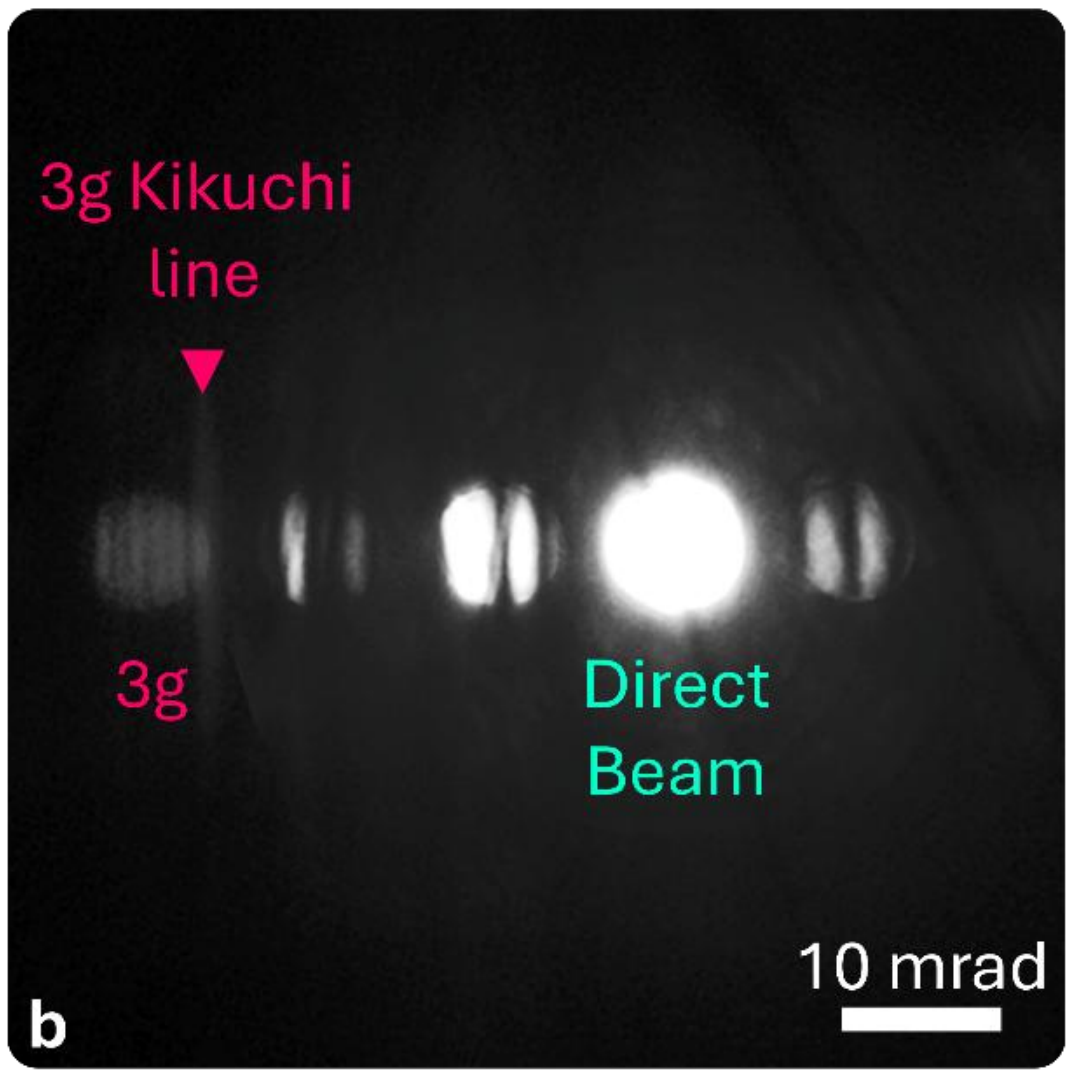

**Fig. S1 Schematic illustration of optical setup and the CBED pattern for the large-loop method.** (a) A suitable objective aperture is inserted to block other discs, except the n**g** and (n+1)**g** discs. For a **g•b** ≠ **0** dislocation, a microscopist could collect two micrographs with inside and outside contrast simultaneously. One micrograph is formed on a STEM-BF detector, and another is formed on a STEM-ADF detector. A corresponding CBED pattern is shown in (b). With a proper objective aperture inserted, the 2g disc in (b) was on the ADF detector and 3g disc was on the BF detector for Fig. S2e-S2h generation.

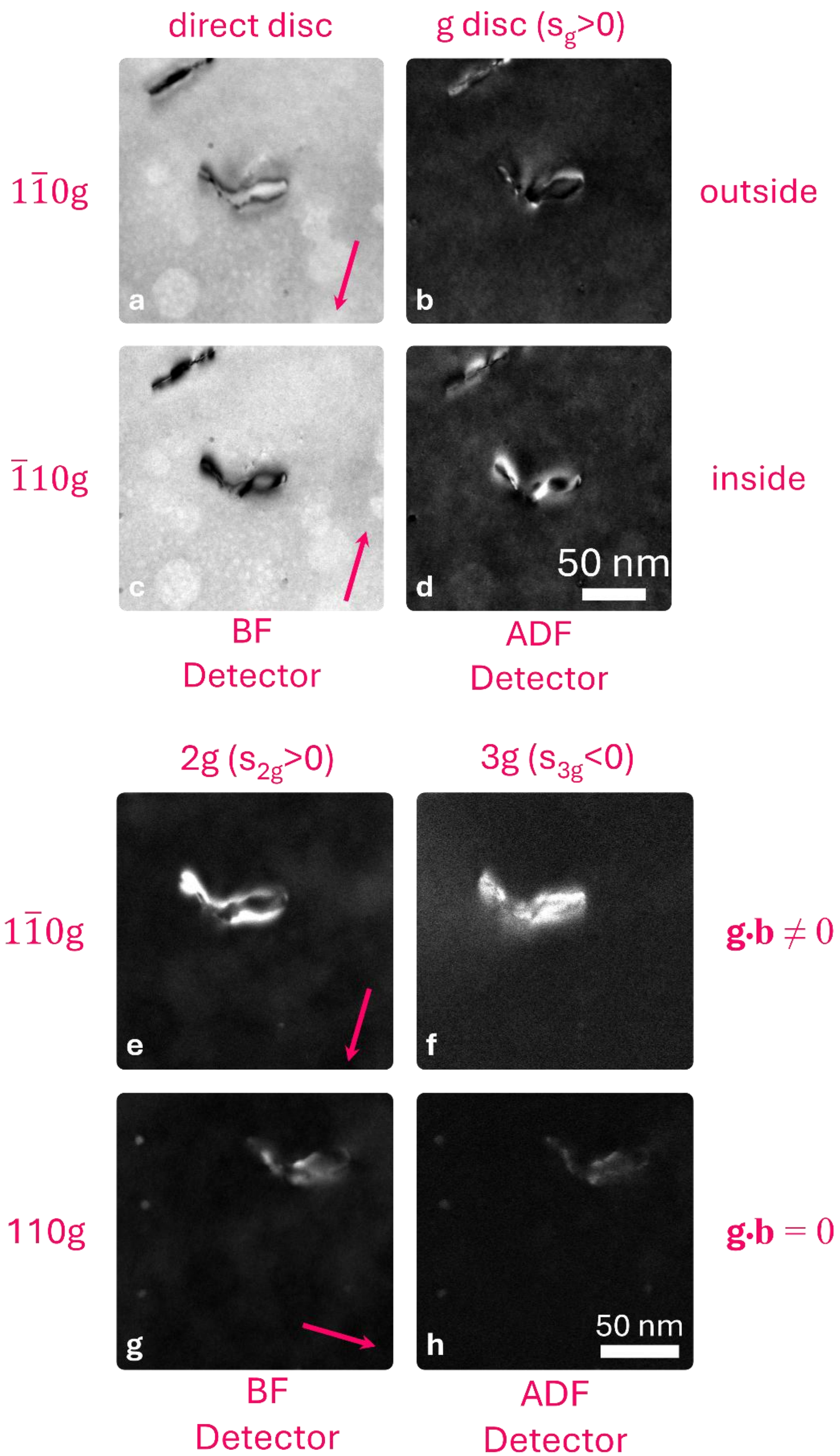

**Fig. S2 A ½<111> loop was imaged for validation.** The inside-outside contrast of the ½<111> loop was achieved in both STEM-BF (a and c) and STEM-DF (b and d) by setting the excitation error of $\mathbf{1\bar{1}0g}$ and $\mathbf{\bar{1}10g}$ positive,

respectively. To demonstrate the superiority of our method for large loops, a ½ <111> loop was imaged under the diffraction condition in Fig. S1b. (e) and (f) show the transition from outside to inside contrast but (g) and (h) show that the loop yielded residual contrast. Note that no tilting in our method.

Another advantage of the method for large loops is the ability to determine the extra plane of an edge-type dislocation line. When the electron beam interacts with the region encompassing the core of an edge dislocation, the dislocation image appears on one side of its physical position in one micrograph but on the opposite in the paired micrograph. As paired micrographs are generated simultaneously, the information within two paired pixels corresponds to the identical location within the TEM foil. By accurately aligning the two paired micrographs, it becomes possible to determine on which side of the dislocation core the dislocation image appears in each micrograph. With the knowledge of its Burgers vector, the extra plane could be determined.

## Large tilt range for well-focused beams

In experiments, 8° tilt is not always the optimal condition. To test the validity of our method in a large range, we simulated the beam shape in multiple positions. By keeping the condition of **110g** systematic row excited and $s_{\mathbf{110g}} = s_{\mathbf{-1-10g}}$, we tilted the sample from 0° to 12° away the 001 zone axis. The simulated probe shapes are shown in Fig. S3. The general probe shapes are consistent under all conditions except the on-zone conditions and the FWHM are nearly identical for all tilted cases at a given depth.

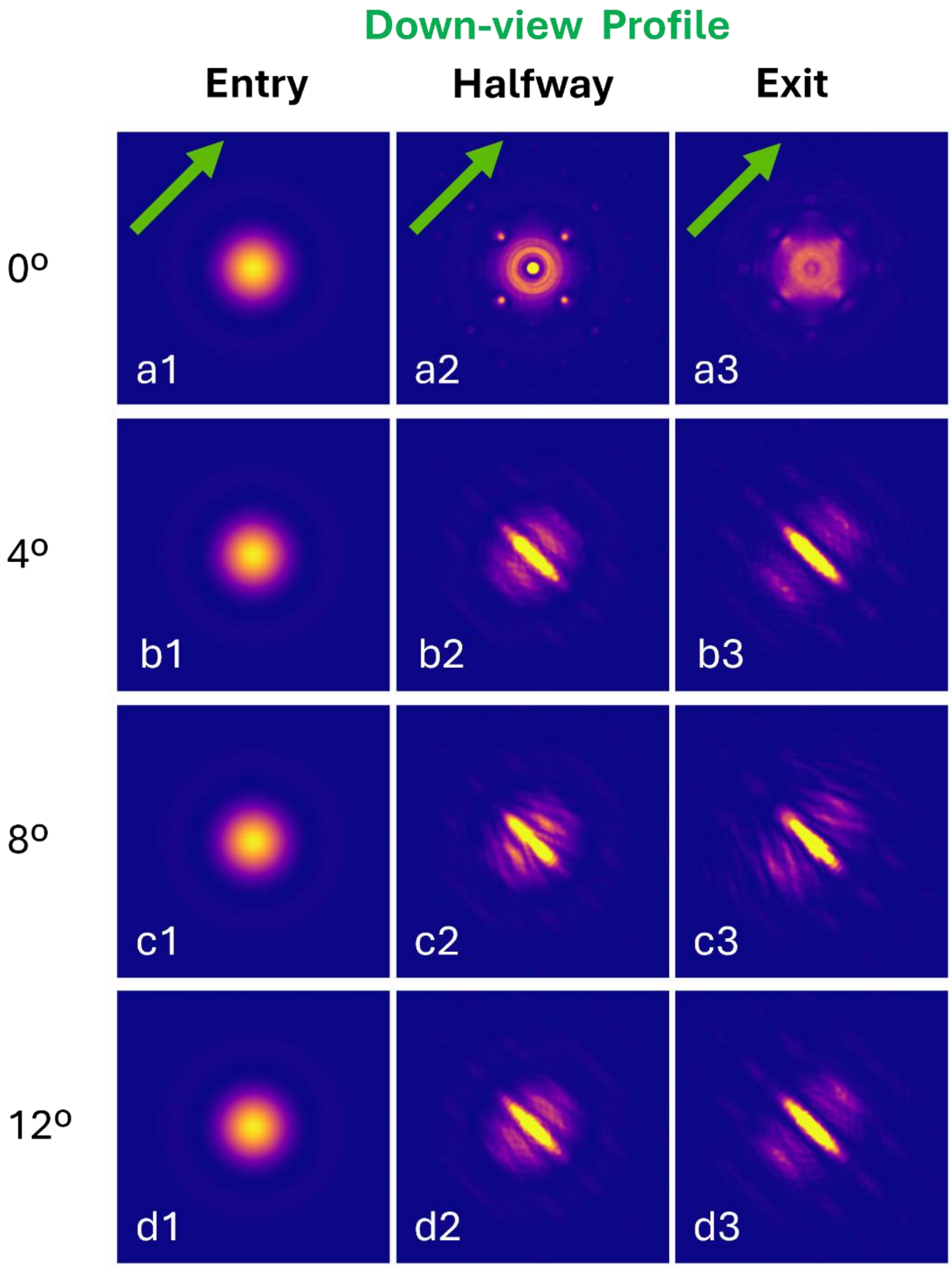

**Fig. S3 Simulation of electron probe in our proposed method in a large tilting range.** The shapes of electron probe under multiple diffraction conditions near the 001 zone axis in BCC Fe. The sample was tilted 0° (a), 4°(b), 8°(c), and 12°(d) away from the zone axis with 110g systematic row excited. The green arrow denotes [110] crystal direction.

## The choice between first and higher order g vectors

Since the diffracted beam intensities depend on the TEM foil thickness, high order **g** imaging requires very thin TEM foils in CTEM mode. Fortunately, the thickness requirement can be significantly relaxed in STEM. In our test, the workable thickness is not smaller than three extinction distance of first order reflections. For iron system imaged with 200 kV STEM whose electron current is ~ 200 pA, the corresponding thickness is ~ 130 nm. With reciprocity theorem, the diffraction theory developed in CTEM remains valid for STEM. Under two-beam dynamical approximation without absorption effects, the diffracted beam intensity in a perfect lattice is $I_g = \frac{1}{1+(s\xi_g)^2} sin^2[\pi \frac{t}{\xi_g}\sqrt{1+(s\xi_g)^2}]$, where $I_g$ is the **g** disc intensity, s is the excitation error of disc **g**, and $\xi_g$ is extinction distance of disc **g**. Due to the sine term, $I_g$ oscillates but the general trend is that the $I_g$ decreases with increasing s, and/or $\xi_g$ (An illustration is Fig. 7.14d (*38*)). For n**g** discs in an identical systematic row, with increasing n, the extinction distance will increase. Thus, the micrographs formed by a higher order **g** disc will be weaker. However, if the disc for imaging is too weak, the background noise will significantly impede analyses. Thus, it is not optimal to always image TEM foils with a high order g vector. To determine the ideal choice on g vector within the proposed methods in this study, we imaged our specimen with **200g** (first order **g**) and **400g** in the **200** systematic row as well as **110g** (first order **g**), and **220g** in the **110g** systematic row, respectively. The results are depicted in Fig. S4. In the **200g** systematic row, the background noise made the **400g** micrograph noisy (Fig. S4b). In contrast, the loop images formed by **200g** were sharp (Fig.S4a). However, when we imaged the loops in the **110g** systematic row, the result reversed. Sharp loop images were recorded within **220g** disc (Fig. S4e). Dislocation loops by **110g** appeared both black, similar to loop images in typical BF micrograph, and bright, similar to loop images in WBDF micrograph (Fig. S4d). If this bright-black image is from a nanosized loop, the loop images become faint, obstructing our Burgers vector determination. One example is highlighted by a white arrow in Fig. S4d, which is fully invisible in Fig. S4c.

The mechanism can be understood due to variation in excitation error induced by the distance differences in reciprocal space. From the insets in Fig. S4, it is clear that the discs in the **200g** row (Fig. S4a and c) are more spaced than the discs in the **110g** row (Fig.S4d and f). When the incident beam is perpendicular to one of these systematic rows, the $s_{2\mathbf{g}}$ would be larger if **g** = **200**. Thus, the micrograph was weaker by the **400g** beam than by **220g**. Another key comparison is between the two micrographs by 1**g** discs (Fig. S4a vs S4d). Fig. S4d is similar to BF. This is due to the relationship between convergence angle and reciprocal distance of **110g** in BCC Fe. Its **110g** is 2.5 $nm^{-1}$ or 6.25 mrad in reciprocal space. When the incident beam was perpendicular to the **110g** systematic row, the **110g** Kikuchi band was evenly divided by the direct disc. In this study, the beam convergence angle is 5 mrad under 200kV. Therefore, the positive (negative) **110g** Kikuchi line stood inside the positive (negative) **110g** disc. Under this diffraction condition, the 'micrograph' by some parts of positive (negative) **110g** was central dark-field images. It is reasonable to form a bright background and a black dislocation in the 'micrograph'. However, other parts in the **110g** disc were with positive excitation error. The 'micrograph' formed by these parts was under weak-beam condition. It is expected to form a dark background and a bright dislocation in the 'micrograph'. The real micrograph recorded by the detector is the sum of the two 'micrographs', resulting in a bright-and-black mixture observed in Fig. S4d. For a microscopist who plans to use all the 1**g** discs for defect imaging, an instant idea to eliminate the mixture is to reduce the convergence angle. However, with decreasing convergence angle, dynamical diffraction effects will be enhanced. It is noteworthy that the situation can change with increasing acceleration voltage of a STEM. At higher voltages, the Ewald's sphere is flatter, and Bragg's angle is reduced. It might be a better option to apply the second order discs in **200g** row and third order discs in **110g** row on a BCC Fe system. Therefore, the choice of what order of diffracted discs for image formation depends on geometrical relationship between the disc and Ewald's sphere.

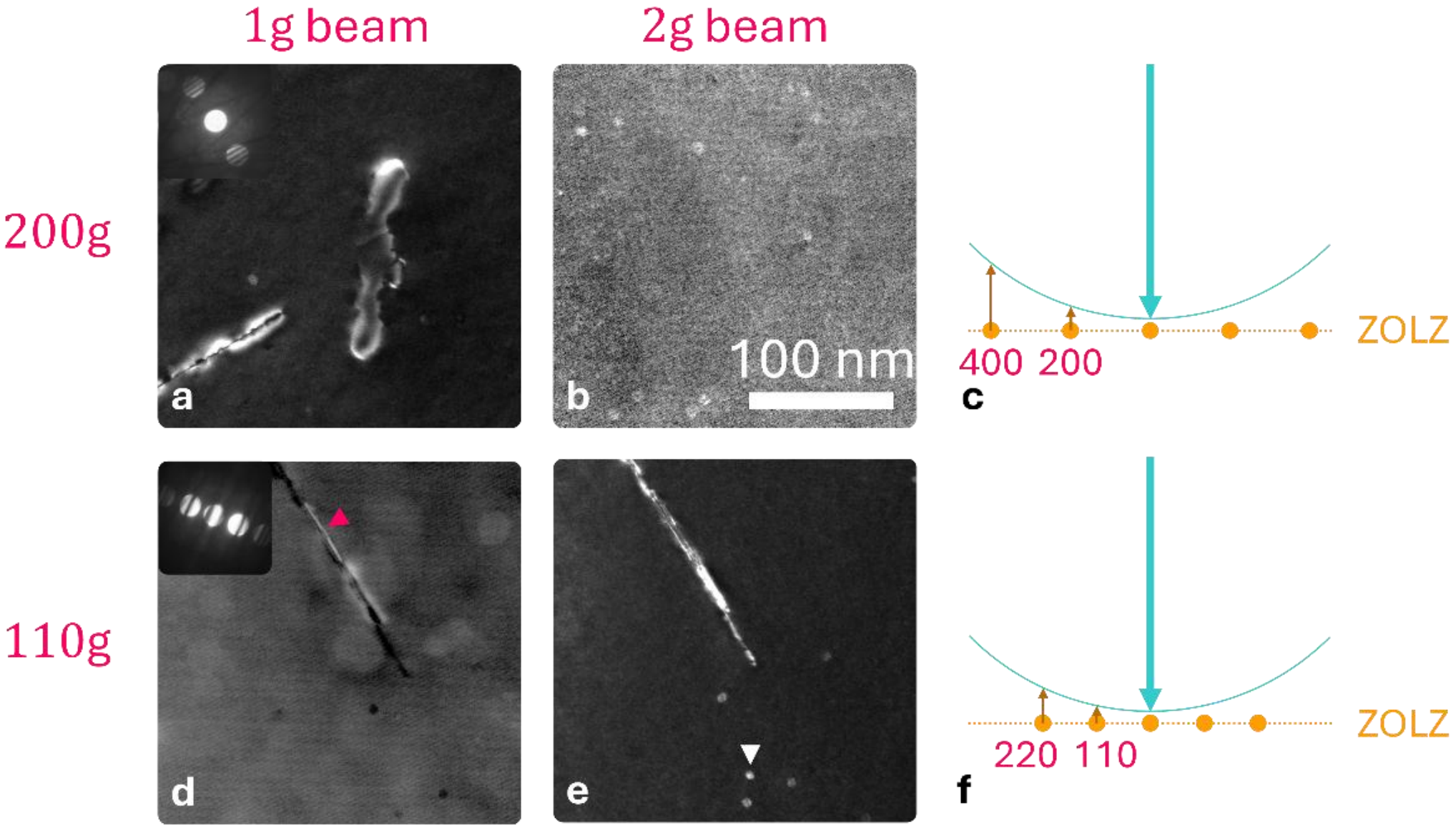

**Fig. S4 A comparison between micrographs formed with first and second order g vectors with method in the main text under [001] zone axis.** The choice between first and second order g vectors depends on the space (reciprocal) between discs in the identical systematic row. The top panel shows the micrographs formed by first order g disc(a) and second order g disc(b) when the **200g** row was excited and the corresponding Ewald's sphere and the zero order Laue zone(c). The bottom panel shows the micrographs formed by first order g disc(d) and second order g disc(e) when the **110g** row was excited and the corresponding Ewald's sphere and the zero order Laue zone (f).

## Image simulation on three different loops

Fig. S5 shows simulated and experimental images from an edge-on [100](200) loop, an inclined [100](111) loop, and an inclined ½[111](111) loop under all four different systematic diffraction rows near 001 zone axis.

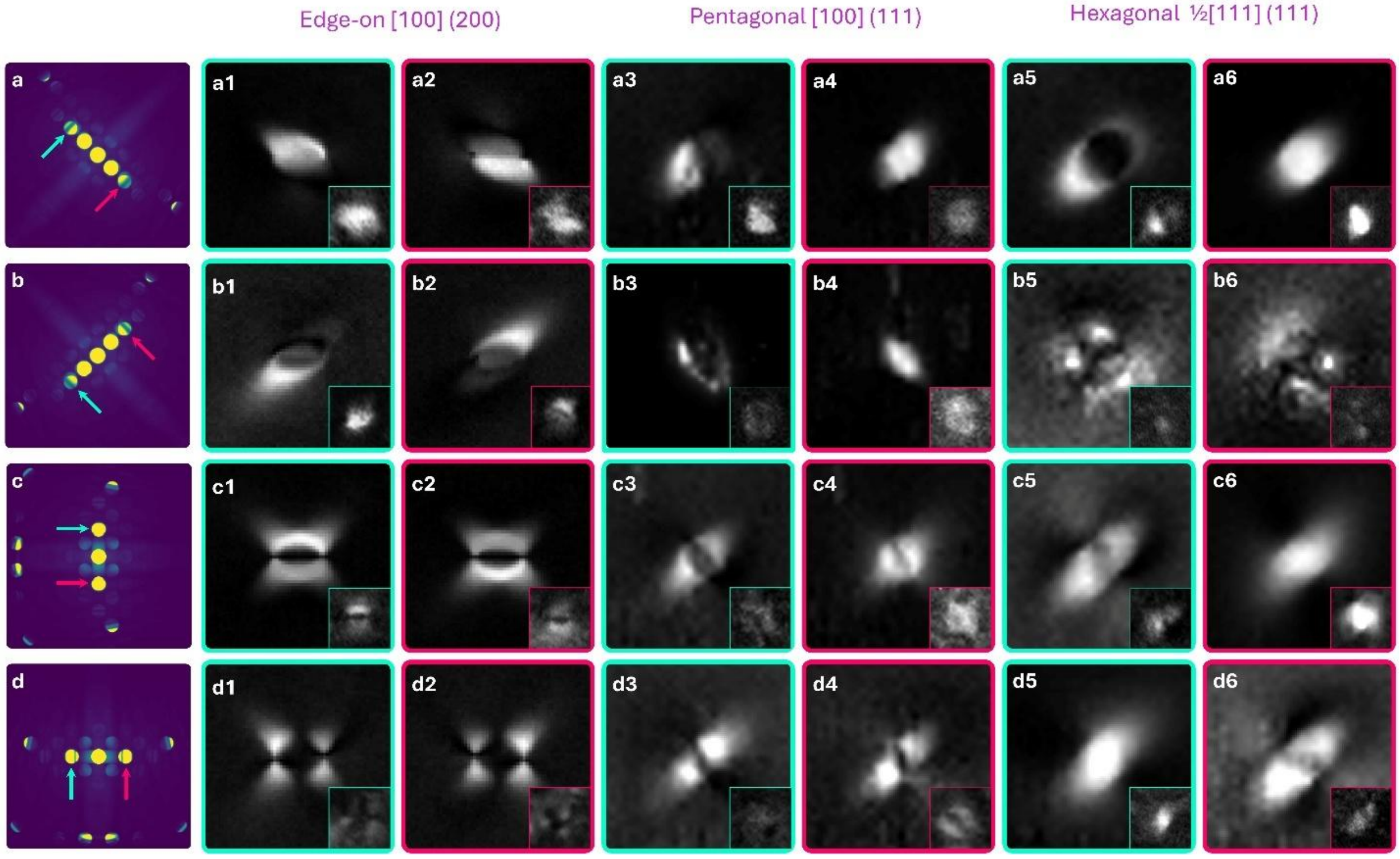

Fig. S5 Simulated images of three different loops (d = 4 nm). The first column is about the diffraction conditions in simulation. The excited systematic reflection rows are **110g**, **1$\bar{1}$0g**, **200g**, and **020g** in sequence from Row **a** to **d. a1**, cyan-framed, is the simulated image formed by the CBED disc pointed by cyan arrow in **a**. **a2**, red-framed, is the simulated image by the CBED disc pointed by red arrow in **a**. **a1** and **a2** are the simulated images from an edge-on **[100]**(200) loop. The inset in **a1**and **a2** are the corresponding experimental images. The identical role applies for **a3** and **a4** from a [100](111) loop, and **a5** and **a6** from a **[111]**(111) loop. All other rows (**b-d**) share the identical logic. B ~ 001.

## Prescreening for 4D-STEM strain mapping

There is another promising application of this new approach: an efficient prescreening tool for time-consuming 4D-STEM strain mapping. We tested our method on petal-shaped dislocation loops (Fig. S6). Our previous study demonstrates that different sections of one petal-shaped loop (the black arrowed loop in Fig. S6a) lie on multiple, adjacent, and parallel atomic planes (*39*), whose internal structure is too fine to be resolved by (g,3g) WBDF (Fig. S6d). Previously, we could only resolve the structure via time-consuming atomic-resolution STEM and 4D-STEM strain mapping (*39*). In contrast, the multi-plane structure can be easily resolved once our new method is applied (Fig. S6e). It is noteworthy that Fig. S6e was recorded at 320kx magnification, contrasting to the atomic-resolution at 5M magnification on our STEM. Thus, with our new method, dozens of strain-inducing defects rigorously parallel to the excited atomic planes in diffraction can be imaged with 0.5-nm resolution in ~ 15 seconds, a typical timespan for one STEM micrograph formation. With ultra-fine details like those in Fig. S6e, subsequent time-consuming 4D-STEM can target the most representative and intriguing defects, improving experiment efficiency.

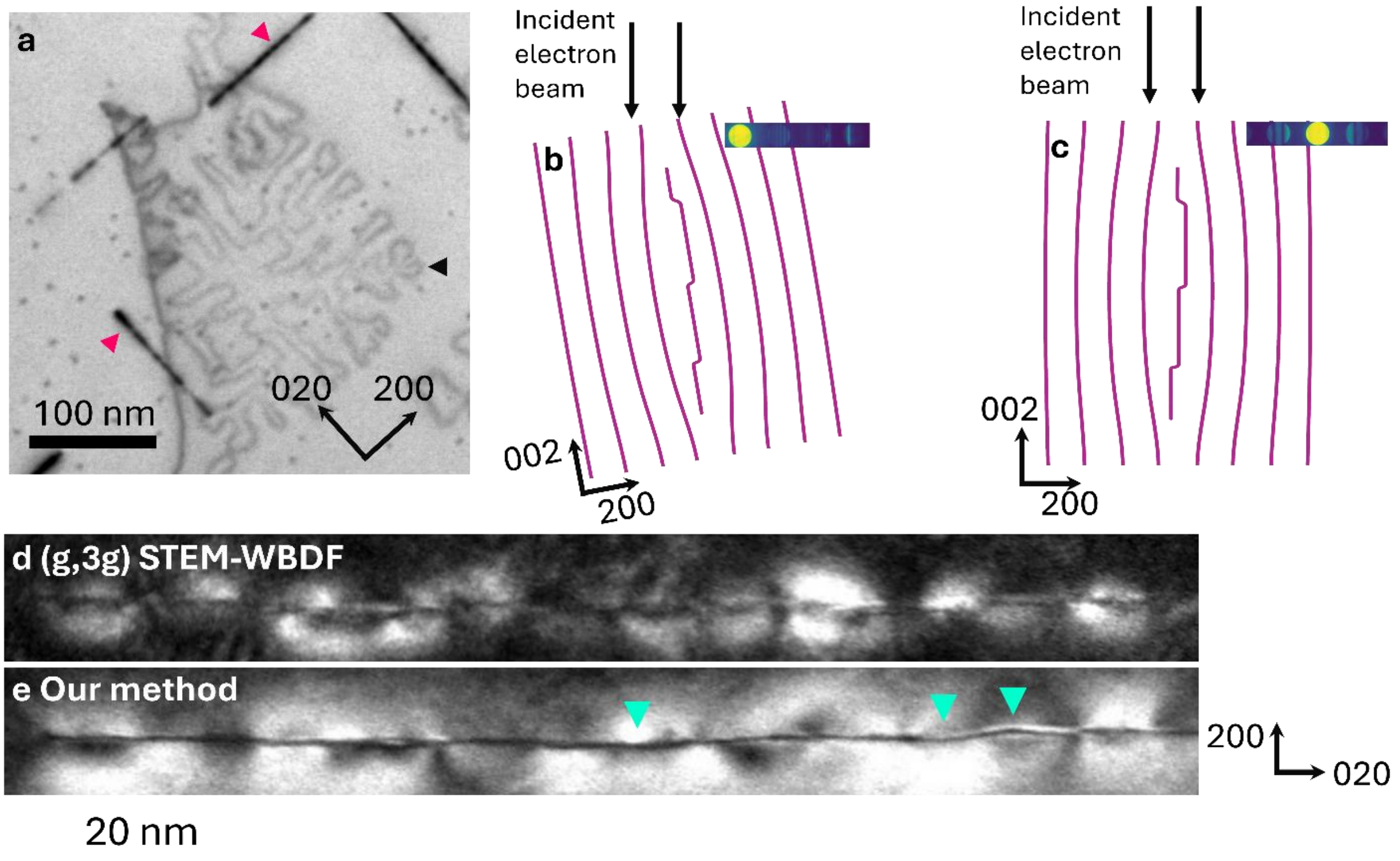

Fig. S6 (a) Petal-shaped loops are recorded using on-zone STEM-BF with B = 001. The specimen is Fe-8Cr (wt%) irradiated to 0.35 dpa at 10-5 dpa/s and 450 °C. The black-arrowed is plan-view petal-shaped loop and the red-arrowed are edge-on petals. In our previous study, we demonstrated that the sections in a petal-shaped loop are on multiple parallel lattice planes (39). The multi-plane structure close to edge-on position and the geometrical relationship between the multi-plane edge-on loops with incident electron beam viewed in classical WBDF in (b) and in our new method (c). Since our method aligns the electron beam parallel to the petal, correctly resolving its multi-plane structure. The corresponding inset is the CBED with ±200g equally excited. (d) Although good resolution is provided by conventional (g,3g) WBDF with 200g, inclined (100) planes, illustrated in (c), hinder the discovery of multi-plane structure illustrated in (c) and (d). Three curved sections (dislocation kink) are highlighted by the cyan arrows which could not be resolved in conventional method(d).